\documentclass[twocolumn,aps,pre,floatfix,superscriptaddress,longbibliography]{revtex4-2}
\pdfoutput=1 %This is for the ArXiv submission only
\usepackage{amsmath,amssymb,eucal,graphicx,float,epstopdf,xparse}
\usepackage{mathtools,xfrac,esint}
\usepackage{epsfig,subfigure}
\usepackage[utf8]{inputenc}
\usepackage{pstricks}
\usepackage[colorlinks=true, urlcolor=blue, anchorcolor=blue, citecolor=blue,filecolor=blue,linkcolor=blue,menucolor=blue]{hyperref}

\begin{document}
\title{Finite-time consensus in a compromise process}

\author{P. L. Krapivsky}
\email{pkrapivsky@gmail.com}
\affiliation{Department of Physics, Boston University, Boston, Massachusetts 02215, USA}
\affiliation{Santa Fe Institute, Santa Fe, New Mexico 87501, USA}

\author{A. Yu. Plakhov}
\affiliation{Center for R{\&}D in Mathematics and Applications, Department of Mathematics, University of Aveiro, Aveiro 3810-193, Portugal}
\affiliation{Institute for Information Transmission Problems, Moscow 127051, Russia}

\begin{abstract}
A compromise process describes the evolution of opinions through binary interactions. Opinions are real numbers, and at each step, two randomly selected agents reach a compromise by averaging their pre-interaction opinions. We prove that if the number $N$ of agents is a power of two, then consensus emerges after a finite number of compromise events with probability one; otherwise, consensus cannot be reached in a finite number of steps, provided the initial opinions are in a general position. The number of steps required to reach consensus is random for $N=2^k$ with $k\geq 2$. We prove that the smallest number of steps is $k\cdot 2^{k-1}$ when the initial opinions are in a general position. For $N=4$, we determine the distribution of the number of steps. In particular, we show that it has a purely exponential tail and compute all cumulants.
\end{abstract}
\maketitle

\section{Introduction}
\label{sec:intro}

In a compromise process, $N$ agents are endowed with opinions that are real numbers. In each step, two distinct agents selected at random average their opinions. Thus
\begin{equation}
\label{def:averaging}
(x_i,x_j) \to \left(\frac{x_i+x_j}{2},\frac{x_i+x_j}{2}\right)
\end{equation}
in a compromise event between agents $i$ and $j$ with $i\ne j$ and opinions $x_i$ and $x_j$.

The simplest compromise process \eqref{def:averaging} can be naturally extended to simple graphs (undirected graphs without loops and multiple edges \cite{Diestel}), where averaging events occur only between agents at neighboring vertices. In this context, the process \eqref{def:averaging} becomes a compromise process on the complete graph $K_N$. Compromise processes often exhibit peculiar behaviors on infinite graphs, e.g., hypercubic lattices; for example, the rate of convergence to consensus may change from exponential to algebraic. Compromise processes are also known as averaging or repeated averaging processes \cite{DeGroot,Olshevsky11,Aldous12, BK21,Diaconis22,Sau23,Cao23,Movassagh,Sau24,Sau25,Peres25}.

Compromise processes have been extended to hypergraphs with averaging events involving agents on vertices of hyperedges \cite{hyper-Renaud,hyper-bounded,hyper-Spiro,hyper-3} and simplicial complexes \cite{simplex,simplex-arXiv}; extensions to infinite graphs \cite{inf-graph} and infinite hypergraphs are also straightforward. Another popular extension of the compromise processes allows averaging events only between sufficiently close opinions \cite{Rainer,Deffuant,BKR,Haggstrom,linguistics,hyper-D}. Compromise processes have been applied to modelling social dynamics \cite{Fortunato_RMP}, communication algorithms \cite{multi_13}, linguistics \cite{linguistics}, gossip \cite{gossip}, energy management in smart grids \cite{grid}, etc. Processes with partial averaging have been investigated in the physics literature as toy models of granular systems with inelastic collisions \cite{BK00,Italy_02,Ernst_02, BK03,Garzo_07,Rajesh_17}.

In the simplest compromise process, all opinions quickly converge to the mean
\begin{equation}
\label{mean}
a = \frac{1}{N} \sum_{j=1}^N  x_j
\end{equation}
of all $x_j$. The rate of convergence exhibits rich behaviors explored, e.g., in Refs.~\cite{Diaconis22,Sau23,Cao23,Movassagh,Peres25}.

In this article, we analyze the possibility of {\em complete} convergence, namely, the emergence of consensus after a {\em finite} number of compromise events. To appreciate this remarkable property, let us look at systems with up to four agents. We find
\begin{enumerate}
\item $N=1$. The initial state is consensus.
\item $N=2$. Consensus is reached after one step.
\item $N=3$. Consensus is never reached.
\item $N=4$. Consensus is reached.
\end{enumerate}

The first two assertions are obvious. The third is intuitively plausible and follows from Theorem 1, which asserts that consensus cannot be reached in a finite number of steps when $N$ is not a power of two. More precisely, this surely happens when initial opinions are in a general position. For degenerate initial distributions, it could be possible to reach consensus after a finite number of steps even when $N\ne 2^k$. For $N=3$, the equidistant distribution $(x_1,x_2,x_3)=(a-d, a, a+d)$ is an example of a degenerate initial distribution, and if the first averaging step involves the 1st and 3rd agents, the consensus is reached. Two other possible first steps lead to an opinion distribution identical to that after the first step if the initial opinions are in a general position; in this situation, consensus cannot be reached after a finite number of steps.

Whenever consensus is reached  after a finite number of steps, the route to consensus is non-trivial and often exceedingly long, except for the system with $N=2$. For the degenerate initial distribution of opinions, such as $(x_1,x_2,x_3,x_4)=(a-d, a, a, a+d)$ for $N=4$, the consensus could be reached after one event if it involves the 1st and 4th agents; for the degenerate initial distribution $(x_1,x_2,x_3,x_4)=(a-d, a-c, a+c, a+d)$, the consensus could be reached after two events. If initial opinions are in a general position in the system with $N=4$ agents, the minimal number of compromise events is four, while the average number is $\frac{89}{2}$ as we show below.

The initial opinion distribution $(x_1,\ldots,x_N)$ is in general position, or non-degenerate, if the equation
\begin{equation}
\sum_{j=1}^N n_j x_j=0, \quad  \sum_{j=1}^N n_j =0
\end{equation}
with $n_j\in \mathbb{Z}$ has only trivial solution $n_1=\ldots=n_N=0$. We always assume that the initial opinion distribution is in general position if not stated otherwise.

Our basic results about the possibility of reaching the consensus in a finite number of steps are the following

\bigskip\noindent
{\bf Theorem 1}. (i) When $N=2^k$, the consensus is reached with probability one after a finite number of compromise events for {\em any} initial opinion distribution. (ii) The consensus cannot be reached in a finite number of compromise events for an initial opinion distribution in general position if $N\ne 2^k$.

\bigskip
When $N=2$, consensus is reached after one step. (Recall that {\em different} agents are selected in each compromise event, and initial opinions are in general position, that means $x_1\ne x_2$ when $N=2$.) When $N=2^k$ with $k\geq 2$, the number of steps is finite, but it varies from realization to realization. What is the minimal number of steps? The answer is given by

\bigskip\noindent
{\bf Theorem 2}. If $N=2^k$, the smallest number of steps to reach consensus is equal to $k\cdot 2^{k-1}$.

\bigskip
We prove Theorem 1 in Sec.~\ref{sec:Proof} and Theorem 2 in Sec.~\ref{sec:proof2}. In Sec.~\ref{sec:time}, we discuss more detailed properties of the consensus time distribution, i.e., the number of compromise events leading to consensus, for the systems with $N=2^k$ agents. In Sec.~\ref{sec:N=4}, we consider the model with $N=4$, the only non-trivial system for which we have understood the consensus time distribution in considerable detail. In Sec.~\ref{sec:1-2}, we analyze a randomized version in which the outcome of each compromise event is slightly perturbed from the average, with perturbations continuously distributed and chosen randomly. Consensus never occurs independently of $N$: The system consists of singlets and dublets, and we determine the equilibrium probability distribution. In Sec.~\ref{sec:future}, we discuss the compromise process on arbitrary connected graphs, not necessarily complete graphs. We show that the finite-time consensus is reached with probability one on all subgraphs of the complete graph $K_{2^k}$ containing the $k$ dimensional cube $C_k$ treated as a graph with $V(C_k)=2^k$ vertices and $E(C_k)=k\cdot 2^{k-1}$ edges.

\section{Proof of Theorem 1}
\label{sec:Proof}

We know that consensus is reached for $N=2$. Let us prove that consensus occurs for $N=4$. The initial state
\begin{subequations}
\begin{equation}
\label{4:0}
 [1]+[1]+[1]+[1]
\end{equation}
turns into
\begin{equation}
\label{4:1}
 [1]+[1]+[2]
\end{equation}
after the first averaging event. Hereinafter $[j]$ denotes a cluster of size $j$, that is, $j$ agents with identical opinions.

A compromise event $[1]+[2]\to [2]+[1]$ does not change the structure of the state \eqref{4:1}. The compromise event $[1]+[1]\to [2]$ changes \eqref{4:1} into
\begin{equation}
\label{4:2}
 [2]+[2]
\end{equation}
\end{subequations}
The state $[1]+[3]$ cannot be reached. It is also impossible to return back to the initial state \eqref{4:0}. The system may oscillate between \eqref{4:1} and \eqref{4:2}, but eventually it reaches the consensus via
\begin{equation}
\label{short-4}
[1]+[1]+[2]\to [2]+[2]\to [1]+[2]+[1]\to [4]
\end{equation}
The state $[1]+[2]+[1]$ that has emerged from $[2]+[2]$ is different from the generic $[1]+[1]+[2]$ state, namely $[1]+[2]+[1]$ has equidistant opinions: If $a$ is the opinion of the pair of agents, isolated agents have opinions $a\pm d$.

In the general case $N=2^k$, it may happen that in the initial series of $N/2$ steps, the agents 1 and 2, 3 and 4, $\ldots$, $N-1$ and $N$ successively reach a compromise, thereby forming $N/2$ pairs $[2] + [2] + \ldots +[2]$. In the next series of $N/2$ steps, agents 1 and 3, 2 and 4, 5 and 7, 6 and 8, sequentially reach a compromise, resulting in $N/4$ pairs $[4] + [4] + \ldots +[4]$. Continuing the sequence of compromise steps in the same manner, at the $l$-th series of steps with $1 \le l \le k$,\, $N/2^l$ pairs $[2^l] + [2^l] + \ldots +[2^l]$ are obtained, and eventually, after $k$ series of steps, the consensus $[2^k]$ is achieved.
The above procedure takes
\begin{equation}
\label{Mk}
M_k = k\cdot 2^{k-1}
\end{equation}
steps to achieve the consensus.

The probability $p$ of realization of such a sequence of steps, although small, is positive. Therefore, the probability of failure to reach consensus after $M$ steps does not exceed $1-p$, and after $qM$ steps does not exceed the value $(1-p)^q$, which tends to zero as $q \to \infty$.

We now prove absence of consensus for $N \ne 2^k$. Assume that the initial distribution $(x_1,\ldots,x_N)$ is in general position, and note that the mean \eqref{mean} of all $x_j$ is conserved. From the formulae of dynamics one sees that at each step of the process, all agents are of the form
\begin{equation}
x_i^{(t)} = 2^{-t} \sum_{j=1}^N n_{ij} x_j,
\end{equation}
where $t$ is the step number and $n_{ij}$ are nonnegative integers (depending on $t$) satisfying
\begin{equation}
\label{eq}
\sum_{j=1}^N n_{ij} = 2^t.
\end{equation}

If consensus is reached at the step $s$, i.e., $x_i^{(s)} = a$ for all $i$, we have
\begin{equation}
2^{-s} \sum_{j=1}^N n_{ij} x_j = \frac 1N \sum_{j=1}^N  x_j,
\end{equation}
leading to
\begin{equation}
\sum_{j=1}^N m_j x_j = 0 \quad \text{with} \quad m_j = 2^{-s} n_{ij} - \frac{1}{N}\,.
\end{equation}
Using \eqref{eq}, one checks that $\sum_{j=1}^N m_j = 0$. Therefore, due to the non-degeneracy of the initial distribution, all $m_j$ are zero. Thus $n_{ij} = 2^s/N$ implying that $N$ divides $2^s$, in contradiction with the assumption that $N \ne 2^k$.

\section{Proof of Theorem 2}
\label{sec:proof2}

By Eq.~\eqref{Mk}, consensus can be reached in $k\cdot 2^{k-1}$ steps. We now show that it cannot be reached in fewer steps.

Indeed, let us assume that consensus is reached after $m < k\cdot 2^{k-1}$ steps. Each step of the process involves 2 agents, which means that a process of $m$ steps involves $2m$ agents. On average, each agent participates in the process $2m/N$ times (this number may not be an integer). We select an agent that has participated in the process the smallest number of times; without loss of generality, we assume that this is $x_1$, and denote by $l$ the number of its participations in the process,
\begin{equation*}
l \le 2m/N < k.
\end{equation*}

Let the first interaction of the agent $x_1$ occur at the step $t_1$ with the agent $i$; after the interaction its value becomes equal to
$$
x_1^{(t_1)} = \frac{x_1 + x_{i}^{(t_1-1)}}{2} = \frac{x_1}{2} + 2^{-t_1} \sum_{j=1}^N n_{ij} x_j.
$$
Continuing this procedure, we see that after the last interaction of the agent 1, say at the step $t_l$ (note that $t_l \le m$), one obtains
\begin{equation}\label{eq1}
x_1^{(m)} = x_1^{(t_l)} = 2^{-l} x_1 + 2^{-t_l} \sum_{j=1}^N n_{ij} x_j.
\end{equation}
Note that the agent $i$ and the nonnegative integers $n_{ij}$ in these formulae depend on the step number; additionally, the values $n_{ij}$ in Eq.~\eqref{eq1} satisfy the relation
\begin{equation}\label{eq0}
2^{-l} + 2^{-t_l} \sum_{j=1}^N n_{ij} = 1.
\end{equation}

According to our assumption of consensus after $m$ steps, we have
\begin{equation}\label{eq2}
x_1^{(m)} = \frac 1N \sum_{j=1}^N x_j.
\end{equation}
Equating expressions \eqref{eq1} and \eqref{eq2} and recalling that $N = 2^k$, we get
$$
2^{k-l} x_1 + 2^{k-t_l} \sum_{j=1}^N n_{ij} x_j = \sum_{j=1}^N x_j,
$$
whence $\sum_{j=1}^N m_j x_j = 0$ with
$$m_1 =  2^{k-l} + 2^{k-t_l} n_{i1} - 1 > 0$$
and $m_j =  2^{k-t_l} n_{ij} - 1$ for $j \ne 1$.
From formula \eqref{eq0} it follows that $\sum_{j=1}^N m_j = 0$, in contradiction with the assumption that the initial opinion distribution $(x_1, \ldots, x_N)$ is in general position.

As a by-product of the proof, we conclude that in a process with the minimal number of steps, each agent $x_j$ participates in exactly $k$ events.

\section{Consensus time distribution}
\label{sec:time}

In this section, we consider the compromise process involving $N=2^k$ agents. We assume the initial opinion distribution is in a general position. With this assumption, consensus is only possible when $N$ is a power of two.

In each compromise event, we randomly select two distinct agents. We then replace their opinions with the average of the two opinions. The total number of possible compromise events is $\binom{N}{2}$. If the selected agents have equal opinions, nothing changes. Still, for theoretical treatment, it is preferable to allow such events with the same probability as events involving agents with different opinions. The compromise process continues until a consensus is reached.

Denote by $P_k(s)$ the probability to reach consensus in $s$ steps. This probability distribution is trivial for $k=0$ and $k=1$,
\begin{equation}
\label{P01}
P_0(s)=\delta_{s,0}     \quad \text{and} \quad P_1(s)=\delta_{s,1},
\end{equation}
and highly non-trivial for $k\geq 2$. For $k=2$, the analytical treatment is still possible, see Sec.~\ref{sec:N=4}.

Computing $P_k(s)$ with $k\geq 3$ appears too challenging, so one would like to probe more basic features of $P_k(s)$. Here are a few such characteristics:
\begin{itemize}
\item The average number of steps $A_k=\langle s \rangle$ before reaching consensus.
\item The minimal number of steps $m_k$ required to reach consensus.
\item The probability $\mu_k=P_k(m_k)$ to reach consensus in the minimal number of steps.
\end{itemize}

The average number of steps before reaching consensus is given by
\begin{equation}
\label{Ak}
A_k = \sum_{s\geq m_k}s\, P_k(s)
\end{equation}
The distribution $P_k(s)$ must decay faster than $s^{-2}$ to ensure that the sum in \eqref{Ak} converges. In Sec.~\ref{sec:N=4}, we compute the distribution $P_2(s)$ and show, among other features, that it has a purely exponential large $s$ tail.

By Theorem 2, if the initial opinion distribution is in general position, the minimal number of steps equals
\begin{equation}
\label{mk}
m_k = k\cdot 2^{k-1}
\end{equation}
In contrast to $m_k$ known for all $k\geq 0$, we know $\mu_k$ and $A_k$ only for $k\leq 2$. For $k=0$ and $k=1$, we have [cf. Eq.~\eqref{P01}]
\begin{subequations}
\begin{align}
\label{mMA:0}
&m_0=0, \quad \mu_0=1, \quad A_0=0\\
\label{mMA:1}
&m_1=1, \quad \mu_1=1, \quad A_1=1
\end{align}
For $k=2$,
\begin{equation}
\label{mMA:2}
m_2 = 4, \quad \mu_2 = \frac{1}{54}\,, \quad A_2 = \frac{89}{2}
\end{equation}
\end{subequations}
Recall that the shortest path to consensus
\begin{equation}
\label{short-4:0}
 [1]+[1]+[1]+[1] \to [1]+[1]+[2]
\end{equation}
followed by \eqref{short-4} realizes $m_2=4$. The derivation of the predictions for $\mu_2$ and $A_2$ in \eqref{mMA:2} require straightforward but cumbersome computations (Sec.~\ref{sec:N=4}).

As a digression, we note that the dependence of the distance from the consensus
\begin{equation}
\label{Manh}
D_s=|{\bf x}^{(s)}-{\bf a}|=\sum_{j=1}^N |x_j^{(s)}-a|
\end{equation}
exhibits a phase transition \cite{Diaconis22} when the number of steps crosses
\begin{equation}
\label{SN}
S_N=\frac{1}{2}N\log_2 N
\end{equation}
(The choice of $L^1$, equivalently the Manhattan norm in \eqref{Manh}, is important.) The convergence is worst when $N-1$ opinions are equal, ${\bf x}^{(0)}=\{1,0,\ldots,0\}$ without loss of generality, and \eqref{SN} applies to this case. More precisely, when $N\gg 1$, the distance \eqref{Manh} remains very close to the initial $D_0=2-\frac{2}{N}$ distance when $\sigma=s/S_N<1$, and $D_s\to 0$ when $\sigma>1$. Amusingly, $S_N=m_k$ when $N=2^k$. This can be coincidental as $S_N$ applies to all $N$ and implies a small $L^1$ distance from consensus, while $m_k$ applies only when $N$ is a power of two and gives the minimal number of steps to complete consensus. Furthermore, we proved \eqref{mk} when initial opinions are in general position, while the distribution ${\bf x}^{(0)}=\{1,0,\ldots,0\}$ is highly degenerate.

\section{Compromise process when $N=4$}
\label{sec:N=4}

In contrast to the system with two agents, the system with $N=4$ agents is nontrivial, but it remains tractable. To begin, we derive the announced results \eqref{mMA:2} and compute $P(5)$ and $P(6)$. [To avoid cluttering formulas, in this section we write $P(s)$ instead of $P_2(s)$.] Next, we compute the cumulant generating function, which encodes the average $\langle s\rangle$, the variance $\langle\!\langle s^2\rangle\!\rangle\equiv \langle s^2\rangle-\langle s\rangle^2$, and all higher cumulants $\langle\!\langle s^n\rangle\!\rangle$. Finally, we derive the exact formula for $P(s)$, from which we deduce that in the $s\to\infty$ limit, $P(s)$ decreases exponentially.

\subsection{Derivation of \eqref{mMA:2}}
\label{sec:S123}

We already know $m_2=4$. We compute $\mu_2=P(4)$ by noting that the first step \eqref{short-4:0} occurs with probability 1. The consecutive steps in \eqref{short-4} occur  with probabilities 1/6, 2/3, and 1/6. Thus
\begin{equation}
\label{P-4-min}
P(4) = \frac{1}{6}\cdot \frac{2}{3}\cdot \frac{1}{6}= \frac{1}{54}  \approx 0.0185.
\end{equation}
For instance, $[1]+[1]+[2]\to [2]+[2]$ occurs with probability 1/6 since there are $\binom{4}{2}=6$ pairs of agents, and we must choose the $[1]+[1]$ pair.

Let $S_1,S_2,S_3$ be the average number of steps to consensus starting from the following three states:
\begin{equation}
\label{S123:def}
\begin{split}
&S_1:  \quad [1]+[1]+[2]\to  [4]\\
&S_2:  \quad [2]+[2]\to  [4]\\
&S_3:  \quad [1]+[2]+[1]\to  [4]
\end{split}
\end{equation}
A straightforward analysis gives
\begin{equation}
\label{S123:eqs}
\begin{split}
&S_1 =   \tfrac{1}{6}S_2+ \tfrac{5}{6}S_1+1\\
&S_2 =   \tfrac{2}{3}S_3+ \tfrac{1}{3}S_2+1\\
&S_3 =   \tfrac{1}{6}S_3+ \tfrac{2}{3}S_1+1
\end{split}
\end{equation}
from which
\begin{equation}
\label{S:123}
S_1 =   \tfrac{87}{2}, \quad   S_2 =   \tfrac{75}{2}, \quad   S_3 =36
\end{equation}

The first step \eqref{short-4:0} brings the system to $[1]+[1]+[2]$, so the average number of steps  to consensus is $A_2 =S_1+1$. Using \eqref{S:123} we arrive at the announced result $A_2 = \frac{89}{2}$.

\subsection{$P(5)$ and $P(6)$}

To compute $P(5)$ we note that one possibility for an extra step is
\begin{equation}
\label{112}
[1]+[1]+[2]\to [1]+ [1]+[2]
\end{equation}
that occurs with probability $5/6$ followed by \eqref{short-4}. If the compromise occurs between agents with identical opinions, so the net effect is zero. In the $[2]+[2]$ arrangement such irrelevant compromise happens with probability $\frac{1}{3}$ and in $[1]+[2]+[1]$ it happens with probability $\frac{1}{6}$. Thus
\begin{equation*}
\frac{P(5)}{P(4)} = \frac{5}{6}+\frac{1}{3}+\frac{1}{6} =  \frac{4}{3}
\end{equation*}
implying that consensus is reached in five steps with probability
\begin{equation}
\label{P-2-5}
P(5) = \frac{2}{81} \approx 0.0247
\end{equation}
exceeding $P(4)$. The probability that consensus is reached in six steps is found from
\begin{equation*}
\tfrac{P(6)}{P(4)} = \left(\tfrac{5}{6}\right)^2+\left(\tfrac{1}{3}\right)^2+\left(\tfrac{1}{6}\right)^2 +\tfrac{5}{6}\cdot \tfrac{1}{3}  +\tfrac{5}{6}\cdot \tfrac{1}{6}+\tfrac{1}{3}\cdot \tfrac{1}{6} = \tfrac{47}{36}
\end{equation*}
to yield
\begin{equation}
\label{P-2-6}
P(6) = \frac{47}{1944} \approx 0.0242.
\end{equation}
The ratio of $P(6)$ to $P(5)$ is $\frac{P(6)}{P(5)} = \frac{47}{48}<1$. Thus $P(5)$ is a local maximum: $P(5) > P(4$ and $P(5) > P(6)$. Moreover, $P(5)$ is a global maximum and $P(s)$ is a decreasing function for $s\geq 5$, viz. $P(s)>P(s+1)$. This observation is supported by simulations, albeit we haven't proved it.

\subsection{Cumulants and the tail}
\label{sec:tail}

In Sec.~\ref{sec:S123}, we computed the average number of steps to consensus, viz., $S_1 = \langle s_1\rangle$, $S_2= \langle s_2\rangle$ and $S_3= \langle s_3\rangle$, starting from the states appearing in \eqref{S123:def}. We now  introduce the cumulant generating functions
\begin{equation}
\sigma_1 = \langle e^{\lambda s_1}\rangle, \quad \sigma_2= \langle e^{\lambda s_2}\rangle, \quad \sigma_3= \langle e^{\lambda s_3}\rangle
\end{equation}
corresponding to the same starting states. These quantities satisfy equations
\begin{equation}
\label{sigma123}
\begin{split}
&e^{-\lambda} \sigma_1 =   \dfrac{1}{6}\sigma_2 + \dfrac{5}{6}\sigma_1 \\
&e^{-\lambda} \sigma_2 =   \dfrac{2}{3}\sigma_3+ \dfrac{1}{3}\sigma_2\\
&e^{-\lambda} \sigma_3 =   \dfrac{1}{6}\sigma_3+ \dfrac{2}{3}\sigma_1+\dfrac{1}{6}
\end{split}
\end{equation}
generalizing Eqs.~\eqref{S123:eqs}.  
Indeed, starting from $[1] + [1] + [2]$ and joining with probability $1/6$ the two agents $[1]$ and $[1]$, one comes to the state $[2] + [2]$. Otherwise (that is, selecting one of the agents $[1]$ and one of the agents in $[2]$, or selecting two agents in $[2]$), with probability $5/6$ one remains in $[1] + [1] + [2]$. Similarly, starting from $[2] + [2]$ and choosing, with probability $1/3$, two agents in one of the pairs $[2]$, one remains in $[2] + [2]$. Otherwise, with probability $2/3$ one passes to $[1] + [2] + [1]$. Finally, if we start from $[1] + [2] + [1]$, with probability $1/6$ one chooses two agents $[1]$ and $[1]$ and immediately comes to $[4]$. Otherwise, with probability $1/6$ one chooses two agents from $[2]$ and then remains in $[1] + [2] + [1]$, and joining one agent $[1]$ and one agent from $[2]$, with probability $2/3$ one passes to $[1] + [1] + [2]$.

Further, the system of linear equations \eqref{sigma123} can be  represented in the matrix form $e^{-\lambda}\boldsymbol{\sigma} = \mathbb{A}\boldsymbol{\sigma} + \boldsymbol{e}_3$, with $\boldsymbol{\sigma} = (\sigma_1, \sigma_2, \sigma_3)^T$,\, $\boldsymbol{e}_3 = (0, 0, 1/3)^T$, and
$$
\mathbb{A} = \left(  \begin{array}{ccc}
5/6 & 1/6 & 0\\ 0 & 1/3 & 2/3\\ 2/3 & 0 & 1/6
\end{array} \right)
$$
from which $\boldsymbol{\sigma} = \big( e^{-\lambda}\,\mathbb{I} - \mathbb{A} \big)^{-1} \boldsymbol{e}_3$, where $\mathbb{I}$ denotes the unit matrix. Inverting the matrix, we find
\begin{equation*}
\label{sigma123:sol}
\sigma_1 =   \frac{2}{D}\,, ~~\sigma_2 =   \frac{12 e^{-\lambda}  - 10}{D}\,, ~~ \sigma_3 =   \frac{5-21 e^{-\lambda}  + 18 e^{-2\lambda}}{D}
\end{equation*}
where $D(\lambda) =  108 e^{-3\lambda} - 144 e^{-2\lambda} +  51 e^{-\lambda}  - 13$.

\begin{figure}[ht]
\begin{center}
\includegraphics[width=0.45\textwidth]{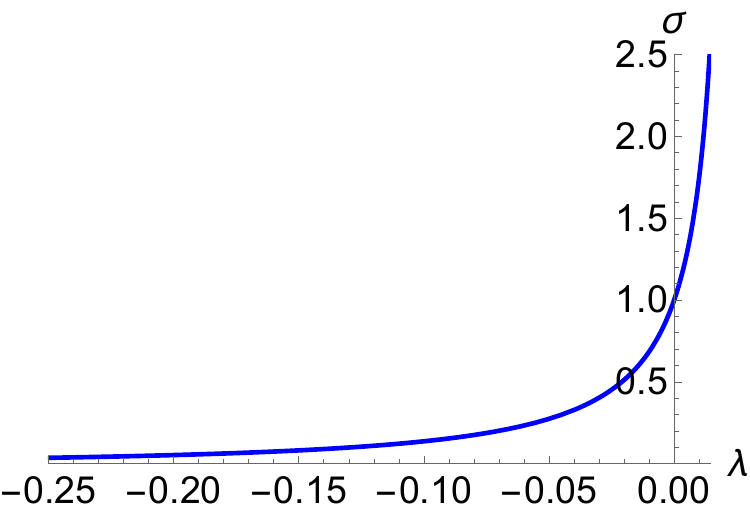}
\caption{The cumulant generating function $\sigma(\lambda)$ is given by \eqref{sigma:sol}. It has a simple pole at $\lambda_c \approx 0.024\,477\,843$. }
\label{Fig:sigma-L}
  \end{center}
\end{figure}

The number of steps starting from the initial state \eqref{4:0} is $s=s_1+1$. The corresponding cumulant generating function $\sigma = \langle e^{\lambda s}\rangle= e^{\lambda}\sigma_1$ is therefore (Fig.~\ref{Fig:sigma-L})
\begin{equation}
\label{sigma:sol}
\sigma =   \frac{2}{108 e^{-4\lambda} - 144 e^{-3\lambda} +  51 e^{-2\lambda}  - 13  e^{-\lambda} }
\end{equation}
Using the standard relation
\begin{equation}
\ln \langle e^{\lambda s}\rangle =
\ln \sigma   = \sum_{n\geq 1} \frac{\lambda^n}{n!}\, \langle\!\langle s^n\rangle\!\rangle
\end{equation}
one can extract the average number of steps $\langle s\rangle=\langle\!\langle s\rangle\!\rangle$, the variance $\langle\!\langle s^2\rangle\!\rangle = \langle s^2\rangle - \langle s\rangle^2$, and all higher cumulants. We recover $A_2=\langle s\rangle=\frac{89}{2}$. The following cumulants read
\begin{equation*}
\begin{split}
\langle\!\langle s^2\rangle\!\rangle & = \frac{6\,675}{4}\\
\langle\!\langle s^3\rangle\!\rangle& = \frac{272\,733}{2}\\
\langle\!\langle s^4\rangle\!\rangle & =  \frac{133\,705\,281}{8}\\
\langle\!\langle s^5\rangle\!\rangle & =  2\,731\,149\,525\\
\langle\!\langle s^6\rangle\!\rangle& =  \frac{2\,231\,527\,901\,505}{4}\\
\langle\!\langle s^7\rangle\!\rangle& = \frac{546\,991\,300\,240\,581}{4}\\
\langle\!\langle s^8\rangle\!\rangle& = \frac{625\,698\,764\,915\,051\,697}{16}\\
\langle\!\langle s^9\rangle\!\rangle& = 2\,780\,920\,968\,559\,151\,285\\
\langle\!\langle s^{10}\rangle\!\rangle& =\frac{18\,797\,127\,943\,068\,382\,118\,685}{4}
\end{split}
\end{equation*}
etc.

We now derive the exact formula for $P(s)$. First, we decompose $\sigma$ given by \eqref{sigma:sol} into simple fractions with respect to $e^\lambda$, and then expand them into series
\begin{eqnarray}
\label{sigma:long}
\sigma &=& e^{4\lambda} \left( \frac{A}{1 - w_0 e^{\lambda}} +  \frac{B_+}{1 - w_+ e^{\lambda}} + \frac{B_-}{1 - w_- e^{\lambda}} \right) \nonumber\\
&=&\sum_{s\geq 0}\left(A\, w_0^s+ B_+\, w_+^s+B_-\, w_-^s\right)e^{\lambda(s+4)}\,.
\end{eqnarray}
Here $w_0 \approx 0.9758$ and $w_\pm \approx 0.1788 \pm 0.3023\,\sqrt{-1}$ are related to the roots of the cubic polynomial in the denominator on the right-hand side in Eq.~\eqref{sigma:sol}; the amplitudes are $A \approx 0.02427$ and $B_\pm \approx -0.002874 \mp 0.003374\,\sqrt{-1}$.
After a simple algebra, one converts \eqref{sigma:long} to
\begin{equation}
\label{sigma:cos}
\sigma = C \sum_{s\ge 4} e^{-\lambda_c s}e^{\lambda s} + C_r \sum_{s\ge 4} e^{-\lambda_r s} \cos(\kappa s - \vartheta) e^{\lambda s}
\end{equation}
with
$C \approx 0.02676$,\, $\lambda_c \approx 0.024477$,\, $C_r \approx 0.5826$,\, $\lambda_r \approx 1.04635$,\, $\kappa \approx 1.0368$, and $\vartheta \approx 0.1403.$ Comparing \eqref{sigma:cos} with the representation for the cumulant generating function
\begin{equation}
\label{sigma-ser}
\sigma = \langle e^{\lambda s} \rangle= \sum_{s\geq 4}  e^{\lambda s} P(s)
\end{equation}
one finally obtains
\begin{equation}
\label{Ps-exact}
P(s) = C\, e^{-\lambda_c s} + C_r\, e^{-\lambda_r s} \cos(\kappa s - \vartheta).
\end{equation}
Since $\lambda_r$ is much greater than $\lambda_c$, the consensus time distribution has an exponential tail,
\begin{equation}
\label{Ps-exp}
P(s) \simeq C\,e^{-\lambda_c s}\quad\text{as}\quad s\to\infty.
\end{equation}
The tail \eqref{Ps-exp} can be alternatively deduced from the closest to the origin simple pole of the cumulant generating function: $\sigma(\lambda)=\frac{C}{\lambda_c-\lambda}+\ldots$.

It is noteworthy that all the probabilities $P(s)$ are rational values. This becomes obvious after realizing that $P(s)$ can be obtained by differentiating $\sigma$ given by \eqref{sigma:sol} $s$ times with respect to $z = e^\lambda$, and then setting $z=0$. This method is also efficient in practical calculations; e.g., one quickly recovers already known $P(s)$ with $s\leq 6$, Eqs.~\eqref {P-4-min}, \eqref{P-2-5}, and \eqref{P-2-6}. The next two probabilities are
\begin{equation*}
P(7) = \frac{133}{5832} \approx 0.028 \quad \text{and} \quad  P(8) = \frac{1537}{69984} \approx 0.020.
\end{equation*}

\section{Distribution of singlets and duplets}
\label{sec:1-2}

In Appendix \ref{ap:states}, we argue that in a compromise process with $N\gg 1$ and $N\ne 2^k$, the majority of {\em observed} configurations are composed of singlets and duplets (pairs of agents with the same opinion). We provided heuristic arguments that the total number of larger groups of agents with identical opinions, i.e., triplets, quadruplets, etc., remains finite in the $N\to\infty$ limit. More precisely, in observed configurations $(1^{n_1} 2^{n_2}3^{n_3} 4^{n_4} 5^{n_5}\cdots)$, the numbers $n_1$ and $n_2$ are asymptotically self-averaging random variables concentrated around $N/2$ and $N/4$, respectively, with typical deviations of order $\sqrt{N}$. The random numbers $n_3$ and $n_4$ remain $O(1)$ when $N\to\infty$. The probability of finding a cluster $[j]$ with $j>4$ decays algebraically with $N$. Thus, each configuration differs from $(1^{n_1} 2^{n_2})$ by a finite number of larger groups, most probably of size three and/or four.

Here, we slightly modify the compromise process. We postulate that the outcome of each compromise event is slightly perturbed from the average. If the perturbations are independently chosen in each event and continuously distributed, an observed configuration is always a composition of singlets and duplets, i.e., it has a form $(1^{N-2m} 2^{m})$, with number $m$ of duplets in the range
\begin{equation}
\label{m-bound}
1\leq m \leq \left\lfloor \frac{N}{2}\right\rfloor
\end{equation}
[Initially, there are no duplets, $m=0$, but after the first compromise event, the system never returns to the initial configuration $(1^N)$.]

Let $p_m(t)$ be the probability to observe $(1^{N-2m} 2^{m})$ at time $t$. In the most natural case of continuous evolution, $m\to m\pm 1$ with rates
\begin{equation}
m\to
\begin{cases}
m+1 & \text{rate} \quad \binom{N-2m}{2}\\
m-1  & \text{rate} \quad 4\binom{m}{2}
\end{cases}
\end{equation}
We set to unity the choice of any pair of distinct agents; the total number of such pairs is $\binom{N}{2}$. For instance, the number of duplets increases, $m\to m + 1$, if two distinct singlets are chosen; the number of such pairs is $\binom{N-2m}{2}$.

The probabilities $p_m(t)$ satisfy the master equation
\begin{eqnarray}
\label{master}
\frac{dp_m}{dt} &=& \binom{N-2m+2}{2}\,p_{m-1} + 4\binom{m+1}{2}\,p_{m+1} \nonumber \\
&-& \left[ \binom{N-2m}{2}+4\binom{m}{2}\right]p_m
\end{eqnarray}

The solution of Eqs.~\eqref{master} subject to the initial condition $p_m(0)=\delta_{m,0}$ is left for future, and here we limit ourselves to the equilibrium or stationary solution describing the probabilities  $p_m(t)$ at $t=\infty$. Equations \eqref{master} become
\begin{eqnarray}
\label{iter}
&&(N-2m+2)(N-2m+1)p_{m-1} + 4m(m+1)p_{m+1} \nonumber \\
&&= [(N-2m)(N-2m-1) + 4m(m-1)]p_m
\end{eqnarray}
Substituting $m=0$ in this formula and taking into account that $p_{-1} =0$, one obtains $p_0 =0$. This fact has an obvious explanation: The configuration $(1^N)$ never appears after the first step.

The probabilities $p_m$ are related by a remarkably simple recurrence
\begin{equation}
\label{rec}
p_{m+1} =\frac{(N -2m)(N -2m -1)}{4m(m+1)}\, p_m
\end{equation}
applicable for $m\geq 1$. Equation \eqref{rec} implies that the probabilities $p_m$ increase with $m$ when $2m$ is smaller than $N/2 -2$ and decrease otherwise. As one could expect, Eq.~\eqref{rec} yields $p_m =0$ for $m \geq \lfloor N/2\rfloor  +1$.

We prove \eqref{rec} by induction. Specializing \eqref{iter} to $m=1$ we obtain
\begin{equation*}
p_2 = \frac{(N-2)(N-3)}{8}\, p_1
\end{equation*}
which agrees with \eqref{rec}. Further, let \eqref{rec} be satisfied for $m-1$, that is,
$$
4m(m-1)\, p_{m}\ =\ (N -2m+2)(N -2m +1)\, p_{m-1}.
$$
Replacing $(N -2m+2)(N -2m +1)\, p_{m-1}$ by $4m(m-1)\, p_{m}$ in formula \eqref{iter}, one obtains \eqref{rec} thereby completing the proof by induction.

Iterating \eqref{rec} yields
\begin{subequations}
\begin{equation}
\label{pmN}
p_m = \frac{C_N}{4^m\, m!(m-1)! (N-2m)!}
\end{equation}
The normalization requirement gives 
\begin{equation*}
C_N\sum_{m=1}^{\lfloor\frac{N}{2}\rfloor}\frac{1}{4^m\, m!(m-1)! (N-2m)!} = 1
\end{equation*}
The sum can be computed exactly allowing us to fix the normalizing factor
\begin{equation}
\label{CN}
C_N = \frac{\sqrt{\pi}\,(N-2)!\, N!}{2^{N-2}\Gamma(N-\frac{1}{2})}
\end{equation}
\end{subequations}
Introducing the auxiliary variable $z = \big( 2m-\frac{N}{2} \big)/\sqrt N$, one derives 
$$
-\ln p_m = (N/4 + \sqrt N z/2)\, \ln 4 + \ln((N/4 + \sqrt N z/2)!) $$ $$+ \ln((N/4 + \sqrt N z/2 - 1)!) + \ln((N/2 - \sqrt N z)!).
$$
Using Stirling's formula $n! \approx n\ln n - n + \frac 12\, \ln(2\pi n)$ one obtains $-\ln p_m \approx 2z^2$, from which one gets an approximate formula
\begin{equation}
p_m \approx \frac{2\sqrt 2}{\sqrt{\pi N}}\ \exp\!\left\{{-\frac{2}{N}\left(2m-\frac{N}{2}\right)^2}\right\}
\end{equation}
valid when $N\gg 1$ and $N/2 - 2m \sim \sqrt N$. Thus, close to the peak, the distribution $p_m$ is asymptotically Gaussian.

The distribution $p_m$ simplifies for the extreme values of $m$. For instance, the minimal number of duplets is observed with probability
\begin{subequations}
\begin{equation}
p_1 = \frac{\sqrt{\pi}\,\,\Gamma(N+1)}{2^N\,\Gamma(N-\frac{1}{2})}\simeq \frac{\sqrt{\pi}\,N^\frac{3}{2}}{2^N}
\end{equation}
For instance, for even $N$, the maximal number of duplets $m=\frac{N}{2}$ is observed with probability
\begin{equation}
\label{m-max}
p_{N/2} = 2^{1-N}\,\frac{\Gamma(N-1)\,\Gamma(\frac{N+1}{2})}{\Gamma(N-\frac{1}{2})\,\Gamma(\frac{N}{2})}\simeq 2^{\frac{1}{2}-N}
\end{equation}
\end{subequations}
In deriving \eqref{m-max}, we have used \eqref{pmN}--\eqref{CN} and simplified the product of gamma functions using the Legendre duplication formula. 

\section{Future directions}
\label{sec:future}

We analyzed the possibility of finite-time consensus on complete graphs $K_N$ and showed that it occurs with probability one when the number of vertices is $N=2^k$ and never occurs for $N\ne 2^k$ (recall that initial opinions are in general position). An intriguing direction of future work is the extension to arbitrary graphs. We consider simple graphs (i.e., no loops and multiple edges), thereby avoiding self-compromise events and ensuring that all possible compromise events between neighboring agents occur at the same rate. We consider only connected graphs; on disconnected graphs, the compromise process proceeds independently on the maximal connected components. In this situation, the number of vertices should be $N=2^k$, although this is just a necessary condition.

The only connected graph with two vertices is $K_2$. For $N=2^k$ with $k\geq 2$, the finite time consensus is impossible on graphs with leaves. Thus, the degree of each vertex is $\geq 2$. The least dense such graph is the ring graph $R_N$. The finite-time consensus occurs on $R_4$. Therefore, for $N=4$, the finite-time consensus occurs on all graphs $R_4\subseteq G\subseteq K_4$. There are three such graphs: $R_4$, $D_4$, and $K_4$, where $D_4$ has an extra `diagonal' edge compared to $R_4$, and thus $R_4\subset D_4\subset K_4$. 

The finite-time consensus does not occur on ring graphs of size $N=2^k$ with $k\geq 3$. It does occur, however, on cubes $C_k$ in $k$ dimensions treated as graphs with $V(C_k)=2^k$ vertices and $E(C_k)=k\cdot 2^{k-1}$ edges. One can prove this assertion by induction. One postulates that the finite-time consensus occurs on $C_{k-1}$. One then treats $C_k$ as composed of the two subgraphs $C_{k-1}$ on the opposite faces, say the floor and the ceiling, with vertical edges between the two $C_{k-1}$. One first reaches a consensus on each $C_{k-1}$; the final consensus is reached by averaging on each vertical edge. We thus arrive at the 

\medskip\noindent
{\bf Conjecture}. For graphs $G$ obeying $C_k\subseteq G\subseteq K_{2^k}$, the consensus is reached with probability one after a finite number of compromise events. The finite-time consensus is never reached for other graphs with $N=2^k$ vertices, and any graphs with $N\ne 2^k$. 

\medskip\noindent
Recall that we deal with initial distributions in general position and consider connected graphs. 

The above conjecture is established only for $k=2$, where $C_2$ is indeed the minimal subgraph of $K_4$, i.e., the finite-time consensus occurs on $C_2$ but is impossible on any of its proper subgraphs $G\subset C_2$. The conjecture asserts that $C_k$ is the unique minimal subgraph of $K_{2^k}$. Proving that $C_k$ is the minimal subgraph and that there are no other minimal subgraphs, or finding more minimal subgraphs when $k\geq 3$, is a challenge for future work. 

For graphs $C_2=R_4$ and $D_4$, one can explore the consensus time distribution in more details. The analysis is the same as for the complete graph $K_4$, Sec.~\ref{sec:N=4}. Here, we merely cite a few results for the graph $C_2$. The analog of the cumulant generating function \eqref{sigma:sol} reads
\begin{equation}
\label{sigma:C_2}
\sum_{s\geq 4} P(s) e^{\lambda s}  =   \frac{1}{32 e^{-4\lambda} - 32 e^{-3\lambda} +  2 e^{-2\lambda}  -  e^{-\lambda} }
\end{equation}
Using \eqref{sigma:C_2}, one finds that the consensus time distribution has an exponential $e^{-\lambda_c s}$ tail with $\lambda_c\approx 0.031715$. One can also extract all the cumulants: the average number of steps $\langle s\rangle=35$,  the variance $\langle\!\langle s^2\rangle\!\rangle = 994$, the third cumulant $\langle\!\langle s^3\rangle\!\rangle = 62694$, etc. One can also deduce the general formula for the probabilities $P(s)$ resembling \eqref{Ps-exact}. For small $s$, one gets
\begin{equation*}
P(4)=P(5)=\frac{1}{32}\,, \quad P(6)=\frac{15}{512}\,, \quad P(7)=\frac{29}{1024}
\end{equation*}

Another extension is to {\em signed} graphs \cite{Harary,Harary65,Zaslavsky,AKR} in which each edge has a positive or negative sign. Signed graphs were introduced \cite{Harary,Harary65} to provide a mathematical basis of Heider's social balance theory \cite{Heider}, and they are popular in social networks \cite{Wasserman,Kleinberg}. The simplest implementation of the compromise process on signed graphs, with averaging events only between agents connected by positive edges, is equivalent to the standard compromise process on graphs obtained from signed graphs by removing edges with negative sign. There could be more interesting versions of the compromise process on signed graphs and generalizations to weighted graphs with weights assigned to edges, and providing rates of compromise events between agents connected to that edge.

\bigskip\noindent
{\bf Data availability statement.} All data that support the findings of this study are included within the article. 

\bigskip\noindent
{\bf Acknowledgments.} PLK is grateful to E. Ben-Naim for the collaboration on the earlier stage of this project.
The work of AYP was supported by CIDMA through FCT, grants UID/4106/2025 and UID/PRR/4106/2025.

\appendix
\section{States}
\label{ap:states}

\begin{figure}
\includegraphics[width=0.29\textwidth]{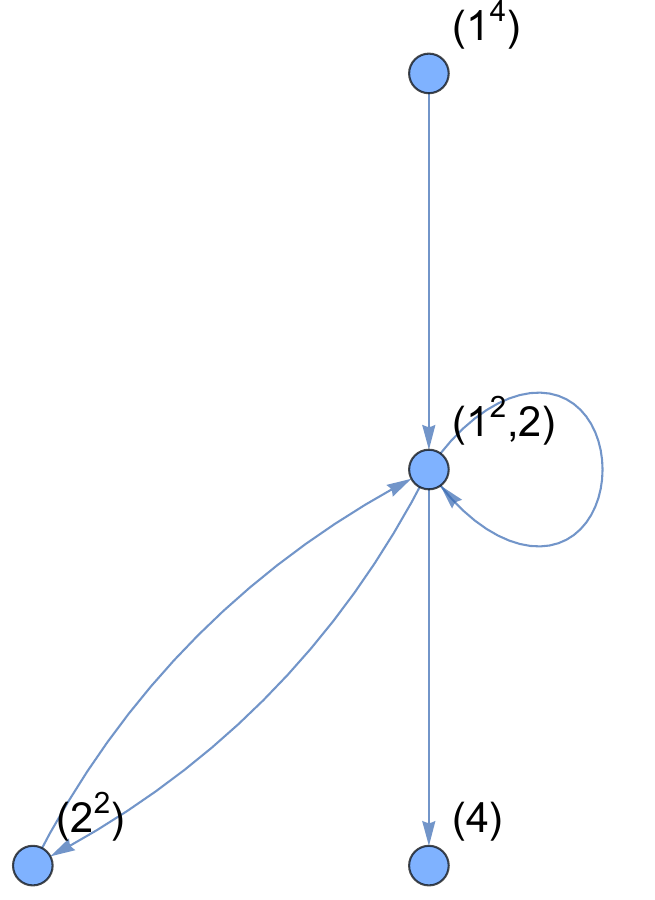}
\includegraphics[width=0.18\textwidth]{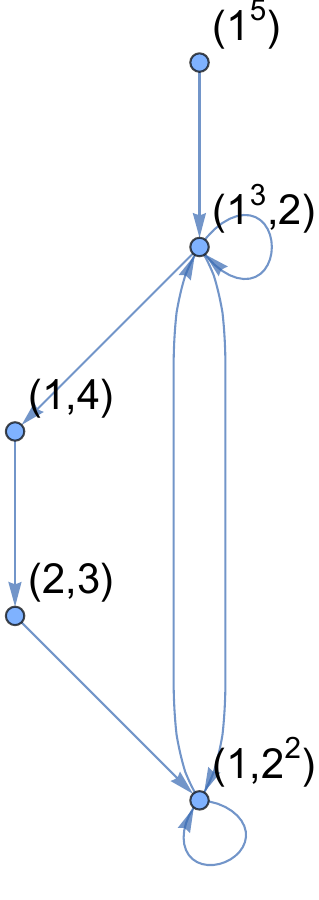}
\caption{The vertices represent the states, and the directed edges represent the transitions between them. Shown are directed graphs for $N=4$ (left) and $N=5$ (right). The system never returns to the initial state, $(1,1,1,1)\equiv (1^4)$ for $N=4$ and $(1^5)$ for $N=5$. The system with $N=4$ eventually reaches the absorbing consensus state $(4)$. The system with $N=5$ wanders ad infinitum on the states $(1^3,2), (1,2^2), (1,4), (2,3)$. }
\label{Fig:45}
\end{figure}

Here, we discuss states that can arise throughout the evolution. A useful perspective is to interpret states as vertices of a graph and transitions as directed edges. For instance, \eqref{short-4:0} is an edge $(1^4)\to (1^2,2)$, where, e.g., $(1^4)$ is the state of four `isolated' agents, i.e., agents with mutually distinct opinions. In Fig.~\ref{Fig:45}, we show the directed graphs for $N=4$ and $N=5$. The representation via the directed graph is incomplete. For $N=4$, for instance, it does not distinguish the generic $(1^2,2)$ states such as the state produced after the first compromise event [see \eqref{4:1}], and the $(1^2,2)$ states that can lead to consensus [see \eqref{short-4}]. It still remains a useful representation, and here we compute the number of vertices $s(N)$ in such directed graphs.

\begin{figure}
\includegraphics[width=0.24\textwidth]{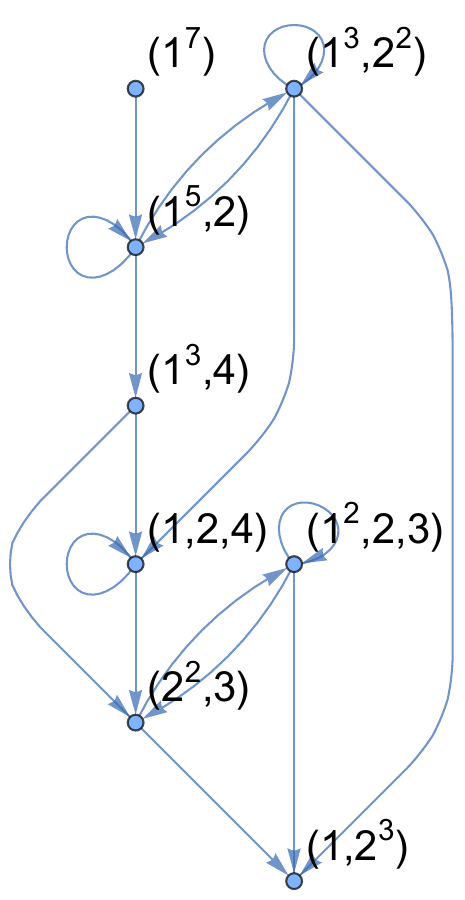}
\caption{The directed graph for $N=7$. The representation via the directed graph is incomplete, i.e., only special states of the type $(1^5, 2)$ can generate $(1^3, 4)$.}
\label{Fig:7}
\end{figure}

We begin with the most basic characteristic, the total number of states $s(N)$. For $N=1$, the only state is the initial state $(1)$, so $s(1)=1$. For $N=2$, the initial state $(1,1)$ gives birth to the consensus state $(2)$; thus, $s(2)=2$. For $N=3$, the initial state $(1^3)$ gives birth to the $(1,2)$, so $s(3)=2$. We also know $s(4)=4$ and $s(5)=5$, see Fig.~\ref{Fig:45}. For $N=6$, there are $s(6)=7$ states:
\begin{equation}
\label{6-states}
(1^6); ~  (1^4, 2); ~ (1^2, 2^2); ~ (1^2, 4); ~ (1, 2, 3); ~(2,4); ~ (2^3)
\end{equation}
For $N=7$, there are $s(7)=8$ states (see also Fig.~\ref{Fig:7}):
\begin{equation}
\label{7-states}
\begin{split}
& (1^7); ~ (1^5, 2); ~  (1^3, 2^2); ~ (1^3, 4); ~ (1^2, 2, 3); \\
&  (1, 2, 4); ~(1^2, 2^3); ~ (1, 2^3); ~ (2^2; 3)
\end{split}
\end{equation}
For $N=8$, there are $s(8)=14$ states:
\begin{equation}
\label{8-states}
\begin{split}
& (1^8); ~ (1^6, 2); ~  (1^4, 2^2); ~ (1^4, 4); ~ (1^3, 2, 3); \\
&  (1^2, 2, 4); ~(1^2, 2^3); ~(1^2, 6); ~ (1, 2^2, 3); ~ (1, 2, 5); \\
& (2^4); ~ (2^2; 4); ~ (2, 3^2); ~ (4^2)
\end{split}
\end{equation}

In Table~\ref{Table:16}, we present the number of states $s(N)$ and the number of partitions $p(N)$. We checked the numbers $s(N)$ appearing in  Table~\ref{Table:16} for $N\leq 8$. The Encyclopedia of integer sequences \cite{sloane25} contains only one sequence beginning with these numbers: the sequence A324753 representing transitive integer partitions. Assuming this alignment is not a coincidental and that $s(N)$ coincides with the sequence A324753, we then used it to obtain the numbers for $N>8$ shown in Table~\ref{Table:16}.

\begin{table}
\vskip 0.5 cm
\renewcommand{\arraystretch}{1.5}{
\begin{tabular}{| c | c | c | c | c | c | c | c | c | c | c | c | c | c | c | c | c |}
\hline
$N$               &  1   &   2   &   3   &  4   &  5  &  6   &   7    &  8     &  9   &  10   & 11   & 12 & 13   & 14  & 15 & 16\\
\hline
$s(N)$          &   1   &   2   &   2   &  4   &  5  &  7   &  8     & 14    & 16  &  23   & 29  & 40  & 49  &  66  & 81 & 109  \\
\hline
$p(N)$          &  1   &   2   &   3    &  5   &  7  &  11 &  15   & 22    &  30  & 42  & 56  & 77 & 101 & 135  & 176 & 231\\
\hline
\end{tabular}
}
\caption{The number of states $s(N)$ and the number of partitions $p(N)$ for  $N\leq 16$.}
\label{Table:16}
\end{table}

The study of partitions goes back to Euler \cite{Euler:book}. One his result is the beautiful expression of the generating function encoding the sequence $p(n)$ through a neat infinite product
\begin{equation}
\label{Euler}
\sum_{n\geq 0}p(n)\,q^n = \prod_{k\geq 1}\frac{1}{1-q^k}
\end{equation}
Using \eqref{Euler}, Euler deduced the large $n$ asymptotic: $\ln p(n)\simeq \pi\sqrt{2n/3}$. Hardy and Ramanujan  derived a more precise asymptotic formula
\begin{equation}
\label{HR}
p(n)\simeq \frac{1}{4\sqrt{3}\,n}\,\exp\!\left[\pi\sqrt{\frac{2n}{3}}\right]
\end{equation}
Rademacher improved \eqref{HR} and derived an {\em exact} formula for $p(n)$; see \cite{Apostol} for a pedagogical derivation. Partitions of integers frequently appear in combinatorics, number theory, and group representations \cite{Andrews,Flajolet,Tableaux,Romik}.

Every state is a partition. Therefore $s(N)\leq p(N)$. The equality holds only for $N=1,2$. For all $N\geq 3$, there are more partitions than states: $s(N)<p(N)$. Looking at Figs.~\ref{Fig:45} and \ref{Fig:7} one notices that every state $1^{n_1} \ldots p^{n_p}$, the largest $p$ such that $n_p>0$ satisfies $p\leq N^{(2)}\equiv 2^{\lfloor \log_2(N)\rfloor}$. This leads to the upper bound
\begin{equation}
\label{bound}
s(N)\leq p(N) - \sum_{j=1}^{N-N^{(2)}} p(j)
\end{equation}
If $N$ is not a power of two, i.e., $N^{(2)}<N$, the bound \eqref{bound} is non-trivial. Using the Hardy-Ramanujan asymptotic formula \eqref{HR} and replacing the sum on the right-hand side (RHS) of \eqref{bound} by the integral, one finds
\begin{eqnarray}
\label{HR-s}
s(N) &\leq& \frac{1}{4\sqrt{3}\,N}\,\exp\!\left[\pi\sqrt{\frac{2N}{3}}\right] \nonumber\\
&-&  \frac{1}{\pi \sqrt{4(N-N^{(2)})}}\,\exp\!\left[\pi\sqrt{\frac{2(N-N^{(2)})}{3}}\right]
\end{eqnarray}
for $N\gg 1$. The first term on the RHS of \eqref{HR-s}, i.e., $p(N)$, is asymptotically dominant. If $N$ is a power of two, the bound \eqref{bound} is trivial: $s(N)\leq p(N)$. The numbers $s(N)$ and $p(N)$ appearing in Table~\ref{Table:16} show that $s(N)<p(N)$ for all $N\geq 3$, including all $N=2^k$ with $k\geq 2$.

The states $1^{n_1} 2^{n_2}$ with $n_1+2n_2=N$ are all present. The number of such states provides the lower bound:
\begin{equation}
\label{low-bound}
s(N)\geq \left\lfloor \frac{N}{2}\right\rfloor + 1
\end{equation}

The states $1^{n_1} 2^{n_2}$ constitute a tiny fraction of all possible states, but the majority of states {\em observed} in the compromise process with a large number of agents are such states. More precisely,
\begin{subequations}
\begin{equation}
\label{n12}
n_1 = \frac{N}{2} + \xi_1, \qquad
n_2 = \frac{N}{4} + \xi_2
\end{equation}
where $\xi_1$ and $\xi_2$ are random with $|\xi_1| \sim |\xi_2| \sim \sqrt{N}$. The number $N-n_1-2n_2$ of `complex' states is a positive random variable that could be of the order of $\sqrt{N}$, similarly to $\xi_1$ and $\xi_2$. Below, we argue that the number of complex states remains finite when $N\to\infty$:
\begin{equation}
\label{Nn12}
N-n_1-2n_2 = O(1)
\end{equation}
\end{subequations}

To appreciate the behaviors \eqref{n12} when $N\gg 1$, we {\em assume} that the number of states different from $1^{n_1} 2^{n_2}$ is much smaller than $N$. Let $N_1=\langle n_1\rangle$ and $N_2=\langle n_2\rangle$. The former quantity obeys a rate equation
\begin{equation}
\label{N12}
\frac{dN_1}{dt} = -2\binom{N_1}{2}+2\cdot 4 \binom{N_2}{2}
\end{equation}
The factor $2$ in the loss term on the RHS of Eq.~\eqref{N12} accounts for two singletons disappearing in the reaction process $[1]+[1]\to [2]$. The factor $2$ in the second term on the RHS accounts for two singletons created in the reaction process $[2]+[2]\to [1]+[2]+[1]$, while the factor $4$ accounts for four possible compromise events. In the steady state, Eq.~\eqref{N12} gives $N_1^2=4N_2^2$ since $N_1\gg 1$ and $N_2\gg 1$. Thus, $N_1=2N_2$.

Similar arguments lead to the rate equation
\begin{equation}
\label{N21}
\frac{dN_2}{dt} = \binom{N_1}{2} - 4 \binom{N_2}{2}
\end{equation}
from which we again deduce $N_1=2N_2$.

A rate equation for $N_4=\langle n_4\rangle$ has a gain term proportional to $N_2$ and the loss term proportional to $N_4 N$. Therefore, in the steady state  $N_4 = O(1)$ which then implies $N_3=O(1)$. Larger groups with $k>4$ agents occur with probabilities vanishing in the $N\to\infty$ limit. For instance, the leading path to $N_6$ and $N_8$ appear to be
\begin{equation*}
\begin{split}
[4]+[4]   & \to [3] + [2] + [3] \to [2] + [4] + [2]\\
             & \to [1] + [6] + [1] \to [8]
\end{split}
\end{equation*}
Writing rate equations for these processes one obtains $N_6\sim N^{-3}$ and $N_8\sim N^{-4}$ which then imply $N_5\sim N^{-3}$ and  $N_7\sim N^{-4}$. The rate equation approach is an uncontrolled approximation for groups of a few agents, so the above estimates for $n_k$ with $k=3,\ldots,8$ are heuristic. Confirming that in a typical realization of the compromise process, there is a finite number of groups of size three and four, $n_3=O(1)$ and $n_4=O(1)$, and $\sum_{k>4}n_k=O(N^{-3})$ of larger groups is an interesting challenge. These estimates constitute a more precise version of the conjecture \eqref{Nn12} which, in conjunction with $N_1=2N_2$, leads to $N_1=N/2$ and $N_2=N/4$ as stated in \eqref{n12}.

\bibliography{references-cons}

\begin{thebibliography}{49}%
\makeatletter
\providecommand \@ifxundefined [1]{%
 \@ifx{#1\undefined}
}%
\providecommand \@ifnum [1]{%
 \ifnum #1\expandafter \@firstoftwo
 \else \expandafter \@secondoftwo
 \fi
}%
\providecommand \@ifx [1]{%
 \ifx #1\expandafter \@firstoftwo
 \else \expandafter \@secondoftwo
 \fi
}%
\providecommand \natexlab [1]{#1}%
\providecommand \enquote  [1]{``#1''}%
\providecommand \bibnamefont  [1]{#1}%
\providecommand \bibfnamefont [1]{#1}%
\providecommand \citenamefont [1]{#1}%
\providecommand \href@noop [0]{\@secondoftwo}%
\providecommand \href [0]{\begingroup \@sanitize@url \@href}%
\providecommand \@href[1]{\@@startlink{#1}\@@href}%
\providecommand \@@href[1]{\endgroup#1\@@endlink}%
\providecommand \@sanitize@url [0]{\catcode `\\12\catcode `\$12\catcode
  `\&12\catcode `\#12\catcode `\^12\catcode `\_12\catcode `\%12\relax}%
\providecommand \@@startlink[1]{}%
\providecommand \@@endlink[0]{}%
\providecommand \url  [0]{\begingroup\@sanitize@url \@url }%
\providecommand \@url [1]{\endgroup\@href {#1}{\urlprefix }}%
\providecommand \urlprefix  [0]{URL }%
\providecommand \Eprint [0]{\href }%
\providecommand \doibase [0]{https://doi.org/}%
\providecommand \selectlanguage [0]{\@gobble}%
\providecommand \bibinfo  [0]{\@secondoftwo}%
\providecommand \bibfield  [0]{\@secondoftwo}%
\providecommand \translation [1]{[#1]}%
\providecommand \BibitemOpen [0]{}%
\providecommand \bibitemStop [0]{}%
\providecommand \bibitemNoStop [0]{.\EOS\space}%
\providecommand \EOS [0]{\spacefactor3000\relax}%
\providecommand \BibitemShut  [1]{\csname bibitem#1\endcsname}%
\let\auto@bib@innerbib\@empty
%</preamble>
\bibitem [{\citenamefont {Diestel}(2017)}]{Diestel}%
  \BibitemOpen
  \bibfield  {author} {\bibinfo {author} {\bibfnamefont {R.}~\bibnamefont
  {Diestel}},\ }\href {https://doi.org/10.1007/978-3-662-53622-3} {\emph
  {\bibinfo {title} {Graph Theory}}}\ (\bibinfo  {publisher} {Springer},\
  \bibinfo {address} {Heidelberg},\ \bibinfo {year} {2017})\BibitemShut
  {NoStop}%
\bibitem [{\citenamefont {DeGroot}(1974)}]{DeGroot}%
  \BibitemOpen
  \bibfield  {author} {\bibinfo {author} {\bibfnamefont {M.~H.}\ \bibnamefont
  {DeGroot}},\ }\bibfield  {title} {\bibinfo {title} {Reaching a consensus},\
  }\href {https://doi.org/10.2307/2285509} {\bibfield  {journal} {\bibinfo
  {journal} {J. Amer. Statist. Association}\ }\textbf {\bibinfo {volume}
  {69}},\ \bibinfo {pages} {118} (\bibinfo {year} {1974})}\BibitemShut
  {NoStop}%
\bibitem [{\citenamefont {Olshevsky}\ and\ \citenamefont
  {Tsitsiklis}(2011)}]{Olshevsky11}%
  \BibitemOpen
  \bibfield  {author} {\bibinfo {author} {\bibfnamefont {A.}~\bibnamefont
  {Olshevsky}}\ and\ \bibinfo {author} {\bibfnamefont {J.~N.}\ \bibnamefont
  {Tsitsiklis}},\ }\bibfield  {title} {\bibinfo {title} {Convergence speed in
  distributed consensus and averaging},\ }\href
  {https://doi.org/10.1137/110837462} {\bibfield  {journal} {\bibinfo
  {journal} {SIAM Review}\ }\textbf {\bibinfo {volume} {53}},\ \bibinfo {pages}
  {747} (\bibinfo {year} {2011})}\BibitemShut {NoStop}%
\bibitem [{\citenamefont {Aldous}\ and\ \citenamefont
  {Lanoue}(2012)}]{Aldous12}%
  \BibitemOpen
  \bibfield  {author} {\bibinfo {author} {\bibfnamefont {D.}~\bibnamefont
  {Aldous}}\ and\ \bibinfo {author} {\bibfnamefont {D.}~\bibnamefont
  {Lanoue}},\ }\bibfield  {title} {\bibinfo {title} {A lecture on the averaging
  process},\ }\href {https://doi.org/10.1214/11-PS184} {\bibfield  {journal}
  {\bibinfo  {journal} {Probab. Surv.}\ }\textbf {\bibinfo {volume} {9}},\
  \bibinfo {pages} {90} (\bibinfo {year} {2012})}\BibitemShut {NoStop}%
\bibitem [{\citenamefont {Ben-Naim}\ and\ \citenamefont
  {Krapivsky}(2021)}]{BK21}%
  \BibitemOpen
  \bibfield  {author} {\bibinfo {author} {\bibfnamefont {E.}~\bibnamefont
  {Ben-Naim}}\ and\ \bibinfo {author} {\bibfnamefont {P.~L.}\ \bibnamefont
  {Krapivsky}},\ }\bibfield  {title} {\bibinfo {title} {Monotonicity in the
  averaging process},\ }\href {https://doi.org/10.1088/1751-8121/ac354f}
  {\bibfield  {journal} {\bibinfo  {journal} {J. Phys. A}\ }\textbf {\bibinfo
  {volume} {54}},\ \bibinfo {pages} {494002} (\bibinfo {year}
  {2021})}\BibitemShut {NoStop}%
\bibitem [{\citenamefont {Chatterjee}\ \emph {et~al.}(2022)\citenamefont
  {Chatterjee}, \citenamefont {Diaconis}, \citenamefont {Sly},\ and\
  \citenamefont {Zhang}}]{Diaconis22}%
  \BibitemOpen
  \bibfield  {author} {\bibinfo {author} {\bibfnamefont {S.}~\bibnamefont
  {Chatterjee}}, \bibinfo {author} {\bibfnamefont {P.}~\bibnamefont
  {Diaconis}}, \bibinfo {author} {\bibfnamefont {A.}~\bibnamefont {Sly}},\ and\
  \bibinfo {author} {\bibfnamefont {L.}~\bibnamefont {Zhang}},\ }\bibfield
  {title} {\bibinfo {title} {A phase transition for repeated averages},\ }\href
  {https://doi.org/10.1214/21-AOP1526} {\bibfield  {journal} {\bibinfo
  {journal} {Ann. Probab.}\ }\textbf {\bibinfo {volume} {50}},\ \bibinfo
  {pages} {1} (\bibinfo {year} {2022})}\BibitemShut {NoStop}%
\bibitem [{\citenamefont {Caputo}\ \emph {et~al.}(2023)\citenamefont {Caputo},
  \citenamefont {Quattropani},\ and\ \citenamefont {Sau}}]{Sau23}%
  \BibitemOpen
  \bibfield  {author} {\bibinfo {author} {\bibfnamefont {P.}~\bibnamefont
  {Caputo}}, \bibinfo {author} {\bibfnamefont {M.}~\bibnamefont
  {Quattropani}},\ and\ \bibinfo {author} {\bibfnamefont {F.}~\bibnamefont
  {Sau}},\ }\bibfield  {title} {\bibinfo {title} {Cutoff for the averaging
  process on the hypercube and complete bipartite graphs},\ }\href
  {https://doi.org/10.1214/23-EJP993} {\bibfield  {journal} {\bibinfo
  {journal} {Electron. J. Probab.}\ }\textbf {\bibinfo {volume} {28}},\
  \bibinfo {pages} {1} (\bibinfo {year} {2023})}\BibitemShut {NoStop}%
\bibitem [{\citenamefont {Cao}(2023)}]{Cao23}%
  \BibitemOpen
  \bibfield  {author} {\bibinfo {author} {\bibfnamefont {F.}~\bibnamefont
  {Cao}},\ }\bibfield  {title} {\bibinfo {title} {Explicit decay rate for the
  {G}ini index in the repeated averaging model},\ }\href
  {https://doi.org/10.1002/mma.8711} {\bibfield  {journal} {\bibinfo  {journal}
  {Math. Meth. Appl. Sci.}\ }\textbf {\bibinfo {volume} {46}},\ \bibinfo
  {pages} {3583} (\bibinfo {year} {2023})}\BibitemShut {NoStop}%
\bibitem [{\citenamefont {Movassagh}\ \emph {et~al.}(2024)\citenamefont
  {Movassagh}, \citenamefont {Szegedy},\ and\ \citenamefont
  {Wang}}]{Movassagh}%
  \BibitemOpen
  \bibfield  {author} {\bibinfo {author} {\bibfnamefont {R.}~\bibnamefont
  {Movassagh}}, \bibinfo {author} {\bibfnamefont {M.}~\bibnamefont {Szegedy}},\
  and\ \bibinfo {author} {\bibfnamefont {G.}~\bibnamefont {Wang}},\ }\bibfield
  {title} {\bibinfo {title} {Repeated averages on graphs},\ }\href
  {https://doi.org/10.1214/24-AAP2050} {\bibfield  {journal} {\bibinfo
  {journal} {Ann. Appl. Probab.}\ }\textbf {\bibinfo {volume} {34}},\ \bibinfo
  {pages} {3781} (\bibinfo {year} {2024})}\BibitemShut {NoStop}%
\bibitem [{\citenamefont {Sau}(2024)}]{Sau24}%
  \BibitemOpen
  \bibfield  {author} {\bibinfo {author} {\bibfnamefont {F.}~\bibnamefont
  {Sau}},\ }\bibfield  {title} {\bibinfo {title} {Concentration and local
  smoothness of the averaging process},\ }\href
  {https://doi.org/10.1214/24-EJP1154} {\bibfield  {journal} {\bibinfo
  {journal} {Electron. J. Probab.}\ }\textbf {\bibinfo {volume} {29}},\
  \bibinfo {pages} {1} (\bibinfo {year} {2024})}\BibitemShut {NoStop}%
\bibitem [{\citenamefont {Sau}(2025)}]{Sau25}%
  \BibitemOpen
  \bibfield  {author} {\bibinfo {author} {\bibfnamefont {F.}~\bibnamefont
  {Sau}},\ }\bibfield  {title} {\bibinfo {title} {Tiny fluctuations of the
  averaging process around its degenerate steady state},\ }\href
  {https://doi.org/10.1214/24-AOP1754} {\bibfield  {journal} {\bibinfo
  {journal} {Ann. Probab.}\ }\textbf {\bibinfo {volume} {53}},\ \bibinfo
  {pages} {1919} (\bibinfo {year} {2025})}\BibitemShut {NoStop}%
\bibitem [{\citenamefont {Elboim}\ \emph {et~al.}(2025)\citenamefont {Elboim},
  \citenamefont {Peres},\ and\ \citenamefont {Peretz}}]{Peres25}%
  \BibitemOpen
  \bibfield  {author} {\bibinfo {author} {\bibfnamefont {D.}~\bibnamefont
  {Elboim}}, \bibinfo {author} {\bibfnamefont {Y.}~\bibnamefont {Peres}},\ and\
  \bibinfo {author} {\bibfnamefont {R.}~\bibnamefont {Peretz}},\ }\bibfield
  {title} {\bibinfo {title} {The edge-averaging process on graphs with random
  initial opinions},\ }\href {https://doi.org/10.1073/pnas.2423947122}
  {\bibfield  {journal} {\bibinfo  {journal} {Proc. Natl. Acad. Sci. U.S.A.}\
  }\textbf {\bibinfo {volume} {122}},\ \bibinfo {pages} {e2423947122} (\bibinfo
  {year} {2025})}\BibitemShut {NoStop}%
\bibitem [{\citenamefont {Neuh\"auser}\ \emph {et~al.}(2020)\citenamefont
  {Neuh\"auser}, \citenamefont {Mellor},\ and\ \citenamefont
  {Lambiotte}}]{hyper-Renaud}%
  \BibitemOpen
  \bibfield  {author} {\bibinfo {author} {\bibfnamefont {L.}~\bibnamefont
  {Neuh\"auser}}, \bibinfo {author} {\bibfnamefont {A.}~\bibnamefont
  {Mellor}},\ and\ \bibinfo {author} {\bibfnamefont {R.}~\bibnamefont
  {Lambiotte}},\ }\bibfield  {title} {\bibinfo {title} {Multibody interactions
  and nonlinear consensus dynamics on networked systems},\ }\href
  {https://doi.org/10.1103/PhysRevE.101.032310} {\bibfield  {journal} {\bibinfo
   {journal} {Phys. Rev. E}\ }\textbf {\bibinfo {volume} {101}},\ \bibinfo
  {pages} {032310} (\bibinfo {year} {2020})}\BibitemShut {NoStop}%
\bibitem [{\citenamefont {Hickok}\ \emph {et~al.}(2022)\citenamefont {Hickok},
  \citenamefont {Kureh}, \citenamefont {Brooks}, \citenamefont {Feng},\ and\
  \citenamefont {Porter}}]{hyper-bounded}%
  \BibitemOpen
  \bibfield  {author} {\bibinfo {author} {\bibfnamefont {A.}~\bibnamefont
  {Hickok}}, \bibinfo {author} {\bibfnamefont {Y.}~\bibnamefont {Kureh}},
  \bibinfo {author} {\bibfnamefont {H.~Z.}\ \bibnamefont {Brooks}}, \bibinfo
  {author} {\bibfnamefont {M.}~\bibnamefont {Feng}},\ and\ \bibinfo {author}
  {\bibfnamefont {M.~A.}\ \bibnamefont {Porter}},\ }\bibfield  {title}
  {\bibinfo {title} {A bounded-confidence model of opinion dynamics on
  hypergraphs},\ }\href {https://doi.org/10.1137/21M1399427} {\bibfield
  {journal} {\bibinfo  {journal} {SIAM J. Appl. Dynamical Systems}\ }\textbf
  {\bibinfo {volume} {21}},\ \bibinfo {pages} {1} (\bibinfo {year}
  {2022})}\BibitemShut {NoStop}%
\bibitem [{\citenamefont {Spiro}(2022)}]{hyper-Spiro}%
  \BibitemOpen
  \bibfield  {author} {\bibinfo {author} {\bibfnamefont {S.}~\bibnamefont
  {Spiro}},\ }\bibfield  {title} {\bibinfo {title} {An averaging process on
  hypergraphs},\ }\href {https://doi.org/10.1017/jpr.2021.67} {\bibfield
  {journal} {\bibinfo  {journal} {J. Appl. Probab.}\ }\textbf {\bibinfo
  {volume} {59}},\ \bibinfo {pages} {495} (\bibinfo {year} {2022})}\BibitemShut
  {NoStop}%
\bibitem [{\citenamefont {Cruciani}\ \emph {et~al.}(2024)\citenamefont
  {Cruciani}, \citenamefont {Giacomelli},\ and\ \citenamefont {Lee}}]{hyper-3}%
  \BibitemOpen
  \bibfield  {author} {\bibinfo {author} {\bibfnamefont {E.}~\bibnamefont
  {Cruciani}}, \bibinfo {author} {\bibfnamefont {E.~L.}\ \bibnamefont
  {Giacomelli}},\ and\ \bibinfo {author} {\bibfnamefont {J.}~\bibnamefont
  {Lee}},\ }\bibfield  {title} {\bibinfo {title} {On the convergence of
  nonlinear averaging dynamics with three-body interactions on hypergraphs},\
  }\href {https://doi.org/10.1137/23M1568338} {\bibfield  {journal} {\bibinfo
  {journal} {SIAM J. Appl. Dynamical Systems}\ }\textbf {\bibinfo {volume}
  {23}},\ \bibinfo {pages} {2364} (\bibinfo {year} {2024})}\BibitemShut
  {NoStop}%
\bibitem [{\citenamefont {DeVille}(2021)}]{simplex}%
  \BibitemOpen
  \bibfield  {author} {\bibinfo {author} {\bibfnamefont {L.}~\bibnamefont
  {DeVille}},\ }\bibfield  {title} {\bibinfo {title} {Consensus on simplicial
  complexes: Results on stability and synchronization},\ }\href
  {https://doi.org/10.1063/5.0037433} {\bibfield  {journal} {\bibinfo
  {journal} {Chaos}\ }\textbf {\bibinfo {volume} {31}},\ \bibinfo {pages}
  {023137} (\bibinfo {year} {2021})}\BibitemShut {NoStop}%
\bibitem [{\citenamefont {Caputo}\ \emph {et~al.}(2024)\citenamefont {Caputo},
  \citenamefont {Quattropani},\ and\ \citenamefont {Sau}}]{simplex-arXiv}%
  \BibitemOpen
  \bibfield  {author} {\bibinfo {author} {\bibfnamefont {P.}~\bibnamefont
  {Caputo}}, \bibinfo {author} {\bibfnamefont {M.}~\bibnamefont
  {Quattropani}},\ and\ \bibinfo {author} {\bibfnamefont {F.}~\bibnamefont
  {Sau}},\ }\bibfield  {title} {\bibinfo {title} {Repeated block averages:
  entropic time and mixing profiles},\ }\href
  {https://arxiv.org/abs/2407.16656} {\bibfield  {journal} {\bibinfo  {journal}
  {arXiv:2407.16656}\ } (\bibinfo {year} {2024})}\BibitemShut {NoStop}%
\bibitem [{\citenamefont {Gantert}\ and\ \citenamefont
  {Vilkas}(2024)}]{inf-graph}%
  \BibitemOpen
  \bibfield  {author} {\bibinfo {author} {\bibfnamefont {N.}~\bibnamefont
  {Gantert}}\ and\ \bibinfo {author} {\bibfnamefont {T.}~\bibnamefont
  {Vilkas}},\ }\bibfield  {title} {\bibinfo {title} {The averaging process on
  infinite graphs},\ }\href {https://arxiv.org/abs/2408.06859} {\bibfield
  {journal} {\bibinfo  {journal} {arXiv:2408.06859}\ } (\bibinfo {year}
  {2024})}\BibitemShut {NoStop}%
\bibitem [{\citenamefont {Hegselmann}\ and\ \citenamefont
  {Krause}(2002)}]{Rainer}%
  \BibitemOpen
  \bibfield  {author} {\bibinfo {author} {\bibfnamefont {R.}~\bibnamefont
  {Hegselmann}}\ and\ \bibinfo {author} {\bibfnamefont {U.}~\bibnamefont
  {Krause}},\ }\bibfield  {title} {\bibinfo {title} {Opinion dynamics and
  bounded confidence: Models, analysis and simulation},\ }\href
  {https://www.jasss.org/5/3/2.html} {\bibfield  {journal} {\bibinfo  {journal}
  {J. Artificial Societies and Social Simulation}\ }\textbf {\bibinfo {volume}
  {5}} (\bibinfo {year} {2002})}\BibitemShut {NoStop}%
\bibitem [{\citenamefont {Weisbuch}\ \emph {et~al.}(2002)\citenamefont
  {Weisbuch}, \citenamefont {Deffuant}, \citenamefont {Amblard},\ and\
  \citenamefont {Nadal}}]{Deffuant}%
  \BibitemOpen
  \bibfield  {author} {\bibinfo {author} {\bibfnamefont {G.}~\bibnamefont
  {Weisbuch}}, \bibinfo {author} {\bibfnamefont {G.}~\bibnamefont {Deffuant}},
  \bibinfo {author} {\bibfnamefont {F.}~\bibnamefont {Amblard}},\ and\ \bibinfo
  {author} {\bibfnamefont {J.-P.}\ \bibnamefont {Nadal}},\ }\bibfield  {title}
  {\bibinfo {title} {Meet, discuss, and segregate!},\ }\href
  {https://doi.org/10.1002/cplx.10031} {\bibfield  {journal} {\bibinfo
  {journal} {Complexity}\ }\textbf {\bibinfo {volume} {7}},\ \bibinfo {pages}
  {55} (\bibinfo {year} {2002})}\BibitemShut {NoStop}%
\bibitem [{\citenamefont {Ben-Naim}\ \emph {et~al.}(2003)\citenamefont
  {Ben-Naim}, \citenamefont {Krapivsky},\ and\ \citenamefont {Redner}}]{BKR}%
  \BibitemOpen
  \bibfield  {author} {\bibinfo {author} {\bibfnamefont {E.}~\bibnamefont
  {Ben-Naim}}, \bibinfo {author} {\bibfnamefont {P.~L.}\ \bibnamefont
  {Krapivsky}},\ and\ \bibinfo {author} {\bibfnamefont {S.}~\bibnamefont
  {Redner}},\ }\bibfield  {title} {\bibinfo {title} {Bifurcation and patterns
  in compromise processes},\ }\href
  {https://doi.org/10.1016/S0167-2789(03)00171-4} {\bibfield  {journal}
  {\bibinfo  {journal} {Physica D}\ }\textbf {\bibinfo {volume} {183}},\
  \bibinfo {pages} {190} (\bibinfo {year} {2003})}\BibitemShut {NoStop}%
\bibitem [{\citenamefont {H\"{a}ggstr\"{o}m}(2012)}]{Haggstrom}%
  \BibitemOpen
  \bibfield  {author} {\bibinfo {author} {\bibfnamefont {A.}~\bibnamefont
  {H\"{a}ggstr\"{o}m}},\ }\bibfield  {title} {\bibinfo {title} {A pairwise
  averaging procedure with application to consensus formation in the {D}effuant
  model},\ }\href {https://doi.org/10.1007/s10440-011-9668-9} {\bibfield
  {journal} {\bibinfo  {journal} {Acta Appl. Math.}\ }\textbf {\bibinfo
  {volume} {119}},\ \bibinfo {pages} {185} (\bibinfo {year}
  {2012})}\BibitemShut {NoStop}%
\bibitem [{\citenamefont {Dong}\ \emph {et~al.}(2016)\citenamefont {Dong},
  \citenamefont {Chen}, \citenamefont {Liang},\ and\ \citenamefont
  {Li}}]{linguistics}%
  \BibitemOpen
  \bibfield  {author} {\bibinfo {author} {\bibfnamefont {Y.}~\bibnamefont
  {Dong}}, \bibinfo {author} {\bibfnamefont {X.}~\bibnamefont {Chen}}, \bibinfo
  {author} {\bibfnamefont {H.}~\bibnamefont {Liang}},\ and\ \bibinfo {author}
  {\bibfnamefont {C.-C.}\ \bibnamefont {Li}},\ }\bibfield  {title} {\bibinfo
  {title} {Dynamics of linguistic opinion formation in bounded confidence
  model},\ }\href
  {https://doi.org/https://doi.org/10.1016/j.inffus.2016.03.001} {\bibfield
  {journal} {\bibinfo  {journal} {Information Fusion}\ }\textbf {\bibinfo
  {volume} {32}},\ \bibinfo {pages} {52} (\bibinfo {year} {2016})}\BibitemShut
  {NoStop}%
\bibitem [{\citenamefont {Schawe}\ and\ \citenamefont
  {Hern\'{a}ndez}(2022)}]{hyper-D}%
  \BibitemOpen
  \bibfield  {author} {\bibinfo {author} {\bibfnamefont {H.}~\bibnamefont
  {Schawe}}\ and\ \bibinfo {author} {\bibfnamefont {L.}~\bibnamefont
  {Hern\'{a}ndez}},\ }\bibfield  {title} {\bibinfo {title} {Higher order
  interactions destroy phase transitions in {D}effuant opinion dynamics
  model},\ }\href {https://doi.org/10.1038/s42005-022-00807-4} {\bibfield
  {journal} {\bibinfo  {journal} {Commun. Phys.}\ }\textbf {\bibinfo {volume}
  {5}},\ \bibinfo {pages} {32} (\bibinfo {year} {2022})}\BibitemShut {NoStop}%
\bibitem [{\citenamefont {Castellano}\ \emph {et~al.}(2009)\citenamefont
  {Castellano}, \citenamefont {Fortunato},\ and\ \citenamefont
  {Loreto}}]{Fortunato_RMP}%
  \BibitemOpen
  \bibfield  {author} {\bibinfo {author} {\bibfnamefont {C.}~\bibnamefont
  {Castellano}}, \bibinfo {author} {\bibfnamefont {S.}~\bibnamefont
  {Fortunato}},\ and\ \bibinfo {author} {\bibfnamefont {V.}~\bibnamefont
  {Loreto}},\ }\bibfield  {title} {\bibinfo {title} {Statistical physics of
  social dynamics},\ }\href {https://doi.org/10.1103/RevModPhys.81.591}
  {\bibfield  {journal} {\bibinfo  {journal} {Rev. Mod. Phys.}\ }\textbf
  {\bibinfo {volume} {81}},\ \bibinfo {pages} {591} (\bibinfo {year}
  {2009})}\BibitemShut {NoStop}%
\bibitem [{\citenamefont {Seyboth}\ \emph {et~al.}(2013)\citenamefont
  {Seyboth}, \citenamefont {Dimarogonas},\ and\ \citenamefont
  {Johansson}}]{multi_13}%
  \BibitemOpen
  \bibfield  {author} {\bibinfo {author} {\bibfnamefont {G.~S.}\ \bibnamefont
  {Seyboth}}, \bibinfo {author} {\bibfnamefont {D.~V.}\ \bibnamefont
  {Dimarogonas}},\ and\ \bibinfo {author} {\bibfnamefont {K.~H.}\ \bibnamefont
  {Johansson}},\ }\bibfield  {title} {\bibinfo {title} {Event-based
  broadcasting for multi-agent average consensus},\ }\href
  {https://doi.org/https://doi.org/10.1016/j.automatica.2012.08.042} {\bibfield
   {journal} {\bibinfo  {journal} {Automatica}\ }\textbf {\bibinfo {volume}
  {49}},\ \bibinfo {pages} {245} (\bibinfo {year} {2013})}\BibitemShut
  {NoStop}%
\bibitem [{\citenamefont {Shah}(2009)}]{gossip}%
  \BibitemOpen
  \bibfield  {author} {\bibinfo {author} {\bibfnamefont {D.}~\bibnamefont
  {Shah}},\ }\bibfield  {title} {\bibinfo {title} {Gossip algorithms},\ }\href
  {https://doi.org/10.1561/1300000014} {\bibfield  {journal} {\bibinfo
  {journal} {Found. Trends Netw.}\ }\textbf {\bibinfo {volume} {3}},\ \bibinfo
  {pages} {1} (\bibinfo {year} {2009})}\BibitemShut {NoStop}%
\bibitem [{\citenamefont {C.~Zhao}\ and\ \citenamefont {Chen}(2017)}]{grid}%
  \BibitemOpen
  \bibfield  {author} {\bibinfo {author} {\bibfnamefont {P.~C.}\ \bibnamefont
  {C.~Zhao}, \bibfnamefont {J.~He}}\ and\ \bibinfo {author} {\bibfnamefont
  {J.}~\bibnamefont {Chen}},\ }\bibfield  {title} {\bibinfo {title}
  {Consensus-based energy management in smart grid with transmission losses and
  directed communication},\ }\href {https://doi.org/10.1109/TSG.2015.2513772}
  {\bibfield  {journal} {\bibinfo  {journal} {IEEE Transactions on Smart Grid}\
  }\textbf {\bibinfo {volume} {8}},\ \bibinfo {pages} {2049} (\bibinfo {year}
  {2017})}\BibitemShut {NoStop}%
\bibitem [{\citenamefont {Ben-Naim}\ and\ \citenamefont
  {Krapivsky}(2000)}]{BK00}%
  \BibitemOpen
  \bibfield  {author} {\bibinfo {author} {\bibfnamefont {E.}~\bibnamefont
  {Ben-Naim}}\ and\ \bibinfo {author} {\bibfnamefont {P.~L.}\ \bibnamefont
  {Krapivsky}},\ }\bibfield  {title} {\bibinfo {title} {Multiscaling in
  inelastic collisions},\ }\href {https://doi.org/10.1103/PhysRevE.61.R5}
  {\bibfield  {journal} {\bibinfo  {journal} {Phys. Rev. E}\ }\textbf {\bibinfo
  {volume} {61}},\ \bibinfo {pages} {R5} (\bibinfo {year} {2000})}\BibitemShut
  {NoStop}%
\bibitem [{\citenamefont {Baldassarri}\ \emph {et~al.}(2002)\citenamefont
  {Baldassarri}, \citenamefont {Marconi},\ and\ \citenamefont
  {Puglisi}}]{Italy_02}%
  \BibitemOpen
  \bibfield  {author} {\bibinfo {author} {\bibfnamefont {A.}~\bibnamefont
  {Baldassarri}}, \bibinfo {author} {\bibfnamefont {U.~M.~B.}\ \bibnamefont
  {Marconi}},\ and\ \bibinfo {author} {\bibfnamefont {A.}~\bibnamefont
  {Puglisi}},\ }\bibfield  {title} {\bibinfo {title} {Influence of correlations
  on the velocity statistics of scalar granular gases},\ }\href
  {https://doi.org/10.1209/epl/i2002-00600-6} {\bibfield  {journal} {\bibinfo
  {journal} {Europhysics Letters}\ }\textbf {\bibinfo {volume} {58}},\ \bibinfo
  {pages} {14} (\bibinfo {year} {2002})}\BibitemShut {NoStop}%
\bibitem [{\citenamefont {Ernst}\ and\ \citenamefont {Brito}(2002)}]{Ernst_02}%
  \BibitemOpen
  \bibfield  {author} {\bibinfo {author} {\bibfnamefont {M.~H.}\ \bibnamefont
  {Ernst}}\ and\ \bibinfo {author} {\bibfnamefont {R.}~\bibnamefont {Brito}},\
  }\bibfield  {title} {\bibinfo {title} {Driven inelastic maxwell models with
  high energy tails},\ }\href {https://doi.org/10.1103/PhysRevE.65.040301}
  {\bibfield  {journal} {\bibinfo  {journal} {Phys. Rev. E}\ }\textbf {\bibinfo
  {volume} {65}},\ \bibinfo {pages} {040301} (\bibinfo {year}
  {2002})}\BibitemShut {NoStop}%
\bibitem [{\citenamefont {Ben-Naim}\ and\ \citenamefont
  {Krapivsky}(2003)}]{BK03}%
  \BibitemOpen
  \bibfield  {author} {\bibinfo {author} {\bibfnamefont {E.}~\bibnamefont
  {Ben-Naim}}\ and\ \bibinfo {author} {\bibfnamefont {P.~L.}\ \bibnamefont
  {Krapivsky}},\ }\bibfield  {title} {\bibinfo {title} {The inelastic {M}axwell
  model},\ }in\ \href {https://doi.org/10.1007/b12449} {\emph {\bibinfo
  {booktitle} {Granular Gas Dynamics}}},\ \bibinfo {editor} {edited by\
  \bibinfo {editor} {\bibfnamefont {T.}~\bibnamefont {P\"{o}schel}}\ and\
  \bibinfo {editor} {\bibfnamefont {N.}~\bibnamefont {Brilliantov}}}\ (\bibinfo
   {publisher} {Springer},\ \bibinfo {address} {Berlin},\ \bibinfo {year}
  {2003})\ pp.\ \bibinfo {pages} {65--94}\BibitemShut {NoStop}%
\bibitem [{\citenamefont {Garz\'{o}}\ and\ \citenamefont
  {Santos}(2007)}]{Garzo_07}%
  \BibitemOpen
  \bibfield  {author} {\bibinfo {author} {\bibfnamefont {V.}~\bibnamefont
  {Garz\'{o}}}\ and\ \bibinfo {author} {\bibfnamefont {A.}~\bibnamefont
  {Santos}},\ }\bibfield  {title} {\bibinfo {title} {Third and fourth degree
  collisional moments for inelastic {M}axwell models},\ }\href
  {https://doi.org/10.1088/1751-8113/40/50/002} {\bibfield  {journal} {\bibinfo
   {journal} {J. Phys. A}\ }\textbf {\bibinfo {volume} {40}},\ \bibinfo {pages}
  {14927} (\bibinfo {year} {2007})}\BibitemShut {NoStop}%
\bibitem [{\citenamefont {Prasad}\ \emph {et~al.}(2017)\citenamefont {Prasad},
  \citenamefont {Das}, \citenamefont {Sabhapandit},\ and\ \citenamefont
  {Rajesh}}]{Rajesh_17}%
  \BibitemOpen
  \bibfield  {author} {\bibinfo {author} {\bibfnamefont {V.~V.}\ \bibnamefont
  {Prasad}}, \bibinfo {author} {\bibfnamefont {D.}~\bibnamefont {Das}},
  \bibinfo {author} {\bibfnamefont {S.}~\bibnamefont {Sabhapandit}},\ and\
  \bibinfo {author} {\bibfnamefont {R.}~\bibnamefont {Rajesh}},\ }\bibfield
  {title} {\bibinfo {title} {Velocity distribution of a driven inelastic
  one-component {M}axwell gas},\ }\href
  {https://doi.org/10.1103/PhysRevE.95.032909} {\bibfield  {journal} {\bibinfo
  {journal} {Phys. Rev. E}\ }\textbf {\bibinfo {volume} {95}},\ \bibinfo
  {pages} {032909} (\bibinfo {year} {2017})}\BibitemShut {NoStop}%
\bibitem [{\citenamefont {Harary}(1954)}]{Harary}%
  \BibitemOpen
  \bibfield  {author} {\bibinfo {author} {\bibfnamefont {M.~H.}\ \bibnamefont
  {Harary}},\ }\bibfield  {title} {\bibinfo {title} {On the notion of balance
  of a signed graph},\ }\href {https://doi.org/10.1307/mmj/1028989917}
  {\bibfield  {journal} {\bibinfo  {journal} {Michigan Math. J.}\ }\textbf
  {\bibinfo {volume} {2}},\ \bibinfo {pages} {143} (\bibinfo {year}
  {1954})}\BibitemShut {NoStop}%
\bibitem [{\citenamefont {Harary}\ \emph {et~al.}(1965)\citenamefont {Harary},
  \citenamefont {Norman},\ and\ \citenamefont {Cartwright}}]{Harary65}%
  \BibitemOpen
  \bibfield  {author} {\bibinfo {author} {\bibfnamefont {F.}~\bibnamefont
  {Harary}}, \bibinfo {author} {\bibfnamefont {R.~Z.}\ \bibnamefont {Norman}},\
  and\ \bibinfo {author} {\bibfnamefont {D.}~\bibnamefont {Cartwright}},\
  }\href@noop {} {\emph {\bibinfo {title} {Structural Models}}}\ (\bibinfo
  {publisher} {J. Wiley \& Sons},\ \bibinfo {address} {New York},\ \bibinfo
  {year} {1965})\BibitemShut {NoStop}%
\bibitem [{\citenamefont {Zaslavsky}(1982)}]{Zaslavsky}%
  \BibitemOpen
  \bibfield  {author} {\bibinfo {author} {\bibfnamefont {T.}~\bibnamefont
  {Zaslavsky}},\ }\bibfield  {title} {\bibinfo {title} {Signed graphs},\ }\href
  {https://doi.org/https://doi.org/10.1016/0166-218X(82)90033-6} {\bibfield
  {journal} {\bibinfo  {journal} {Discrete Appl. Math.}\ }\textbf {\bibinfo
  {volume} {4}},\ \bibinfo {pages} {47} (\bibinfo {year} {1982})}\BibitemShut
  {NoStop}%
\bibitem [{\citenamefont {Antal}\ \emph {et~al.}(2006)\citenamefont {Antal},
  \citenamefont {Krapivsky},\ and\ \citenamefont {Redner}}]{AKR}%
  \BibitemOpen
  \bibfield  {author} {\bibinfo {author} {\bibfnamefont {T.}~\bibnamefont
  {Antal}}, \bibinfo {author} {\bibfnamefont {P.~L.}\ \bibnamefont
  {Krapivsky}},\ and\ \bibinfo {author} {\bibfnamefont {S.}~\bibnamefont
  {Redner}},\ }\bibfield  {title} {\bibinfo {title} {Social balance on
  networks: The dynamics of friendship and enmity},\ }\href
  {https://doi.org/https://doi.org/10.1016/j.physd.2006.09.028} {\bibfield
  {journal} {\bibinfo  {journal} {Physica D: Nonlinear Phenomena}\ }\textbf
  {\bibinfo {volume} {224}},\ \bibinfo {pages} {130} (\bibinfo {year}
  {2006})}\BibitemShut {NoStop}%
\bibitem [{\citenamefont {Heider}(1958)}]{Heider}%
  \BibitemOpen
  \bibfield  {author} {\bibinfo {author} {\bibfnamefont {F.}~\bibnamefont
  {Heider}},\ }\href@noop {} {\emph {\bibinfo {title} {The psychology of
  interpersonal relations}}}\ (\bibinfo  {publisher} {J. Wiley \& Sons},\
  \bibinfo {address} {New York},\ \bibinfo {year} {1958})\BibitemShut {NoStop}%
\bibitem [{\citenamefont {Wasserman}\ and\ \citenamefont
  {Faust}(1994)}]{Wasserman}%
  \BibitemOpen
  \bibfield  {author} {\bibinfo {author} {\bibfnamefont {S.}~\bibnamefont
  {Wasserman}}\ and\ \bibinfo {author} {\bibfnamefont {K.}~\bibnamefont
  {Faust}},\ }\href {https://doi.org/10.1017/CBO9780511815478} {\emph {\bibinfo
  {title} {Social Network Analysis: Methods and Applications}}}\ (\bibinfo
  {publisher} {Cambridge University Press},\ \bibinfo {address} {Cambridge,
  UK},\ \bibinfo {year} {1994})\BibitemShut {NoStop}%
\bibitem [{\citenamefont {Easley}\ and\ \citenamefont
  {Kleinberg}(2012)}]{Kleinberg}%
  \BibitemOpen
  \bibfield  {author} {\bibinfo {author} {\bibfnamefont {D.}~\bibnamefont
  {Easley}}\ and\ \bibinfo {author} {\bibfnamefont {J.}~\bibnamefont
  {Kleinberg}},\ }\href {https://doi.org/10.1017/CBO9780511761942} {\emph
  {\bibinfo {title} {Networks, Crowds, and Markets: Reasoning about a Highly
  Connected World}}}\ (\bibinfo  {publisher} {Cambridge University Press},\
  \bibinfo {address} {Cambridge, UK},\ \bibinfo {year} {2012})\BibitemShut
  {NoStop}%
\bibitem [{\citenamefont {Sloane}(2025)}]{sloane25}%
  \BibitemOpen
  \bibfield  {author} {\bibinfo {author} {\bibfnamefont {N.~J.~A.}\
  \bibnamefont {Sloane}},\ }\href {https://oeis.org/A324753} {\bibinfo {title}
  {The on-line encyclopedia of integer sequences ({OEIS})}} (\bibinfo {year}
  {2025})\BibitemShut {NoStop}%
\bibitem [{\citenamefont {Euler}(1988)}]{Euler:book}%
  \BibitemOpen
  \bibfield  {author} {\bibinfo {author} {\bibfnamefont {L.}~\bibnamefont
  {Euler}},\ }\href {https://doi.org/10.1007/978-1-4612-1021-4} {\emph
  {\bibinfo {title} {Introduction to Analysis of the Infinite}}}\ (\bibinfo
  {publisher} {Springer},\ \bibinfo {address} {New York, NY},\ \bibinfo {year}
  {1988})\BibitemShut {NoStop}%
\bibitem [{\citenamefont {Apostol}(1990)}]{Apostol}%
  \BibitemOpen
  \bibfield  {author} {\bibinfo {author} {\bibfnamefont {T.~M.}\ \bibnamefont
  {Apostol}},\ }\href {https://doi.org/10.1007/978-1-4612-0999-7} {\emph
  {\bibinfo {title} {Modular Functions and Dirichlet Series in Number
  Theory}}}\ (\bibinfo  {publisher} {Springer},\ \bibinfo {address} {New York,
  NY},\ \bibinfo {year} {1990})\BibitemShut {NoStop}%
\bibitem [{\citenamefont {Andrews}\ and\ \citenamefont
  {Eriksson}(2004)}]{Andrews}%
  \BibitemOpen
  \bibfield  {author} {\bibinfo {author} {\bibfnamefont {G.~E.}\ \bibnamefont
  {Andrews}}\ and\ \bibinfo {author} {\bibfnamefont {K.}~\bibnamefont
  {Eriksson}},\ }\href {https://doi.org/10.1017/CBO9781139167239} {\emph
  {\bibinfo {title} {Integer Partitions}}}\ (\bibinfo  {publisher} {Cambridge
  University Press},\ \bibinfo {address} {Cambridge, UK},\ \bibinfo {year}
  {2004})\BibitemShut {NoStop}%
\bibitem [{\citenamefont {Flajolet}\ and\ \citenamefont
  {Sedgewick}(2009)}]{Flajolet}%
  \BibitemOpen
  \bibfield  {author} {\bibinfo {author} {\bibfnamefont {P.}~\bibnamefont
  {Flajolet}}\ and\ \bibinfo {author} {\bibfnamefont {R.}~\bibnamefont
  {Sedgewick}},\ }\href {https://doi.org/10.1017/CBO9780511801655} {\emph
  {\bibinfo {title} {Analytic Combinatorics}}}\ (\bibinfo  {publisher}
  {Cambridge University Press},\ \bibinfo {address} {Cambridge, UK},\ \bibinfo
  {year} {2009})\BibitemShut {NoStop}%
\bibitem [{\citenamefont {Fulton}(1996)}]{Tableaux}%
  \BibitemOpen
  \bibfield  {author} {\bibinfo {author} {\bibfnamefont {W.}~\bibnamefont
  {Fulton}},\ }\href {https://doi.org/10.1017/CBO9780511626241} {\emph
  {\bibinfo {title} {Young Tableaux, with Applications to Representation Theory
  and Geometry}}}\ (\bibinfo  {publisher} {Cambridge University Press},\
  \bibinfo {address} {Cambridge, UK},\ \bibinfo {year} {1996})\BibitemShut
  {NoStop}%
\bibitem [{\citenamefont {Romik}(2015)}]{Romik}%
  \BibitemOpen
  \bibfield  {author} {\bibinfo {author} {\bibfnamefont {D.}~\bibnamefont
  {Romik}},\ }\href {https://doi.org/10.1017/CBO9781139872003} {\emph {\bibinfo
  {title} {The Surprising Mathematics of Longest Increasing Subsequences}}}\
  (\bibinfo  {publisher} {Cambridge University Press},\ \bibinfo {address}
  {Cambridge, UK},\ \bibinfo {year} {2015})\BibitemShut {NoStop}%
\end{thebibliography}%

\end{document}